\documentclass{article}
\usepackage{log_2024}						
\usepackage{bm}
\usepackage{booktabs}						
\usepackage{multirow}		
\usepackage{amssymb}
\usepackage{amsfonts}						
\usepackage{graphicx}						
\usepackage{duckuments}						

\newcommand{\grad}{\text{grad}}
\newcommand{\curl}{\text{curl}}
\newcommand{\erf}{\text{erf}}
\newcommand{\Rev}[1]{\textcolor{black}{#1}}
\newcommand{\argmin}{\text{arg min}}

\newcommand{\dive}{\text{div}}
\newcommand{\imag}{\text{im}}

\usepackage[numbers,compress,sort]{natbib}	

\title[Decomposing force fields as flows on graphs reconstructed from stochastic trajectories]{Decomposing force fields as flows on graphs reconstructed from stochastic trajectories}

\author[R. Nartallo-Kaluarachchi et al.]{%
Ramón Nartallo-Kaluarachchi\\
Mathematical Institute, University of Oxford\\
Centre for Eudaimonia and Human Flourishing, University of Oxford\\
\email{nartallokalu@maths.ox.ac.uk}
\And
Paul Expert\\
UCL Global Business School for Health, UCL\And
David Beers\\
Department of Mathematics, UCLA
\And
Alexander Strang\\
Department of Statistics, University of California, Berkeley\And
Morten L. Kringelbach\\
Centre for Eudaimonia and Human Flourishing, University of Oxford\\
Department of Psychiatry, University of Oxford\\
Center for Music in the Brain, Aarhus University\And
Renaud Lambiotte\\
Mathematical Institute, University of Oxford\\
The Alan Turing Institute\And
Alain Goriely
\\
Mathematical Institute, University of Oxford}
\begin{document}
\maketitle
\begin{abstract}
Disentangling irreversible and reversible forces from random fluctuations is a challenging problem in the analysis of stochastic trajectories measured from real-world dynamical systems. We present an approach to approximate the dynamics of a stationary Langevin process as a discrete-state Markov process evolving over a graph-representation of phase-space, reconstructed from stochastic trajectories. Next, we utilise the analogy of the Helmholtz-Hodge decomposition of an edge-flow on a contractible simplicial complex with the associated decomposition of a stochastic process into its irreversible and reversible parts. This allows us to decompose our reconstructed flow and to differentiate between the irreversible currents and reversible gradient flows underlying the stochastic trajectories. We validate our approach on a range of solvable and nonlinear systems and apply it to derive insight into the dynamics of flickering red-blood cells and healthy and arrhythmic heartbeats. In particular, we capture the difference in irreversible circulating currents between healthy and passive cells and healthy and arrhythmic heartbeats. Our method breaks new ground at the interface of data-driven approaches to stochastic dynamics and graph signal processing, with the potential for further applications in the analysis of biological experiments and physiological recordings. Finally, it prompts future analysis of the convergence of the Helmholtz-Hodge decomposition in discrete and continuous spaces.
\end{abstract}
\section{\label{sec: Introduction} Introduction}
The analysis of stochastic trajectories emerging from noisy, fluctuating systems, such as those found in biology, has become an important challenge in data science \cite{Friedrich2011complexity}. Such systems exhibit stochastic dynamics and can be well-modelled by Langevin processes, which are stochastic differential equations (SDEs) describing the time-evolution of variables subject to deterministic forces and noisy fluctuations \cite{Kampen1992}. Examples of such systems include climate \cite{Franzke_O’Kane_2017}, neural \cite{harrison2005stochasticneuronal} and heartbeat dynamics \cite{qu2014heart} as well as fluctuations in microscopic biological processes \cite{Gnesotto2018brokendetailedbalance}, financial time-series and Brownian particles \cite{brown1828brownian}. Data from such systems takes the form of noisy trajectories in which drift and diffusive fields are inaccessible prompting the development of a range of methods to infer and approximate dynamics directly from observed trajectories \cite{Friedrich2011complexity,Frishman2020stochasticforce,bruckner2020inferring,garcia2018reconstruction,genkin2021learning,casert2024learning,elbeheiry2015inferencemap,ferretti2020generallangevin,Brückner2019twostate,battle2016brokendetailedbalance,Lynn2021brokendetailedbalance}. A particular interest lies in the analysis of the nonequilibrium, time-irreversible nature of these processes, consistent with the consumption of energy and the dissipation of entropy to their environments \cite{Gnesotto2018brokendetailedbalance,fang2019nonequilibrium}. Existing techniques for studying nonequilibrium dynamics in data include coarse-graining for estimating probability currents \cite{battle2016brokendetailedbalance,Lynn2021brokendetailedbalance}, measuring time-irreversibility \cite{Roldán2021irreversibility}, thermodynamic uncertainty relations \cite{Horowitz2019uncertainty} or the variance-sum rule for measuring entropy production \cite{diterlizzi2024variancesum}.\\\\
Concurrently, topological and graph-theoretic methods for the analysis of large and complex datasets have emerged as a powerful paradigm within data science \cite{carlsson2009topological,carlsson2020topological,Bronstein2017geometric,Edelsbrunner2010comptop}. Such methods have also found applications in the analysis of time-series through techniques like attractor reconstruction \cite{muldoon1993timeseries} and, more recently, graph and topological signal processing \cite{ortega2018gsp,Schaub2021SignalProcessing,Schaub2022SignalProcessing,Barbarossa2020topsig}. Amongst these methods, the Helmholtz-Hodge decomposition (HHD), which decomposes flows into gradients and circulations, has proved to be a valuable tool in the analysis of empirical signals evolving on discrete objects \cite{Lim2020Hodge,Schaub2020RandomWalks,strang2020applications}. Surprisingly, an analogous decomposition to the HHD can also be applied to stationary Langevin processes, decomposing stochastic dynamics into reversible and irreversible parts \cite{strang2020applications,fang2019nonequilibrium,DaCosta_2023}. In an attempt to bridge the HHD of an SDE with that of an associated flow on a graph, we construct a data-driven approach for approximating a Langevin process with a Markov chain constructed entirely from stochastic trajectories. Firstly, we use a maximum-likelihood estimator to reconstruct the transition intensities of a discrete-state, continuous-time Markov chain that approximates the underlying Langevin process. By taking an appropriate ratio of the rates, we define an edge-flow on a triangulation of phase-space and decompose this flow to gain insight into the irreversible and reversible forces driving the dynamics of the trajectories. Our approach is validated with numerical simulations of both solvable and nonlinear systems and then applied to the analysis of flickering red-blood cell (RBC) experiments and healthy and arrhythmic heartbeats. In particular, our approach is able to capture the difference in the prevalence of irreversible, circulating currents between the dynamics of healthy and \textit{adenosine triphosphate}-depleted (ATP) RBCs \cite{diterlizzi2024variancesum,turlier2019unveiling,betz2008rbc} as well as the increased time-irreversibility of healthy heartbeats compared to those with arrhythmia \cite{peng2019nonequilibrium,costa2005heart}.
\section{Methods}
\subsection{\label{sec: HH} The various faces of the Helmholtz-Hodge decomposition}
\begin{figure*}
    \centering
    \includegraphics[width = \textwidth]{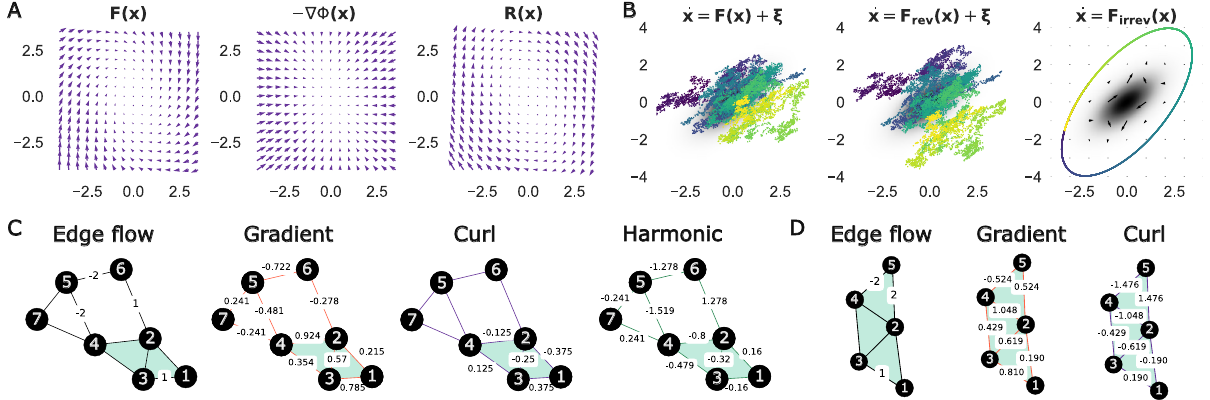}
    \caption{\textbf{The Helmholtz-Hodge decomposition in various spaces.} \textbf{A.} The HHD of a linear stable vector field $\bm{F}$ into a gradient component, $-\nabla \bm{\Phi}$, and a rotational component, $\bm{R}$. \textbf{B.} The HHD of \Rev{the drift field of a} stationary Langevin process into its reversible component, \Rev{the portion of the drift that maintains the stationary distribution (plotted behind - black) as constant despite the noisy fluctuations, and its irreversible component, which drive deterministic rotation around the stationary distribution. The first panel shows a trajectory from the stationary OU process, whilst the second and third show the same trajectory decomposed into its reversible and irreversible dynamics respectively. The arrows in the final panel indicate the stationary probability flux $\bm{J}_{ss}$}. \textbf{C.} The HHD of an edge-flow on a simplicial complex into gradient, curl and harmonic parts. Simplices are oriented naturally according to vertex labels. The gradient part sums to zero around closed loops. The curl part is only defined on triangles (example flow from Ref. \cite{Schaub2020RandomWalks}). \textbf{D.} The HHD of an edge-flow on a triangulation into gradient and curl parts. As the complex has no holes, there is no harmonic flow. Blank edges have zero flow.}
    \label{fig: Helmholtz}
\end{figure*}
The HHD is a special case of the, more general, Hodge decomposition of a flow on a manifold \cite{voisin2010hodge}. Three variants of the HHD include the decomposition of a real-valued vector field \cite{GLOTZL2023127138}, the decomposition of the drift field of stationary Langevin process \cite{DaCosta_2023}, or the decomposition of an edge-flow on a simplicial complex \cite{Lim2020Hodge}. Fig. \ref{fig: Helmholtz} illustrates each of these three faces of the HHD. The Langevin process is described by the SDE,
\begin{align}\label{eq: langevin}
    \dot{\mathbf{x}}(t) & = \bm{F}(\mathbf{x}(t)) + \bm{\xi}(t),
\end{align}
where \Rev{$\bm{F}: \mathbb{R}^{N} \rightarrow  \mathbb{R}^{N}$ is a vector field on $\mathbb{R}^N$} that represents the \Rev{time-constant}, deterministic, driving force of the system, known as the \textit{drift}, whilst $\bm{\xi} \in \mathbb{R}^n$ are \Rev{noise} in the form of Gaussian fluctuations, known as the \textit{diffusion}, with spatially-homogeneous correlations given by $\langle \bm{\xi}(t),\bm{\xi}(t')\rangle = 2\bm{D}\delta(t-t')$ where \Rev{$\bm{D}\in \mathbb{R}^N \times \mathbb{R}^N$} is the constant, positive definite, noise-covariance matrix. \Rev{Alternatively, one can consider the probability density of the process where $P(\mathbf{x},t)$ represents the probability of the process being in state $\mathbf{x}$ at time $t$. The density evolves according to} the Fokker-Planck (FP) equation \cite{Kampen1992},
\begin{align}\label{eq: FP}
    \frac{\partial P(\mathbf{x},t)}{\partial t} = &-\sum_i \frac{\partial}{\partial x_i} \left [ F_iP\right] + \sum_{i,j} \frac{\partial}{\partial x_i \partial x_j} \left[ D_{ij}P\right].
\end{align}
\Rev{This can be written as a conservation of probability law,
\begin{align}
\label{eq: conservation}
    \frac{\partial P}{\partial t}+ \nabla \cdot \bm{J} &=0,
\end{align}
where $\bm{J}(\mathbf{x},t) = \bm{F}(\mathbf{x})P(\mathbf{x},t) - \nabla \cdot (\bm{D}P(\mathbf{x},t))$ represents the probability flux and $\nabla \cdot $ represents the divergence operator (see App. \ref{app: HHD}). The process is described as \textit{stationary} when its density is constant, i.e. when $\partial P / \partial t = 0$, despite the process continuing to evolve in time. At stationarity, we} can decompose the drift as follows \cite{DaCosta_2023},
\begin{align}
    \bm{F}& = \bm{F}_{\text{irr}} + \bm{F}_{\text{rev}},\\
    \bm{F}_{\text{rev}} &= \bm{D}\cdot\nabla \log P_{ss},\label{eq: helmholtzsde}\\
    \bm{F}_{\text{irr}} &=\bm{J}_{ss}/P_{ss} \label{eq: helmholtzsdeirrev},
\end{align}
where $\bm{F}_{\text{irr}}$ \& $\bm{F}_{\text{rev}}$ represent the time-irreversible and time-reversible components of the drift respectively, \Rev{$\nabla$ is the gradient of a scalar potential (see App. \ref{app: HHD}), $P_{ss}(\mathbf{x})$ is the stationary distribution, the solution to Eq. \ref{eq: FP} when $\partial P / \partial t = 0$, and $\bm{J}_{ss}(\mathbf{x})$ is the stationary probability flux, which satisfies $\nabla \cdot \bm{J} = 0$}. More specifically, $\bm{F}_{\text{irr}}$ is odd whilst $\bm{F}_{\text{rev}}$ is even under time-reversal \cite{DaCosta_2023}. The detailed balance condition is satisfied if $\bm{J}_{ss} \equiv \bm{0}$, corresponding to an equilibrium stationary distribution with time-reversible dynamics, and broken if $\bm{J}_{ss} \not\equiv \bm{0}$, corresponding to a nonequilibrium stationary distribution and time-irreversible dynamics \cite{fang2019nonequilibrium}. Thus, information about the (ir)reversibility of the stochastic dynamics can be inferred by decomposing the drift field of the underlying stationary Langevin process. As shown in panel B of Fig. \ref{fig: Helmholtz}, the reversible component of the drift climbs the stationary distribution whilst the diffusion causes random fluctuations. The irreversible part of the drift drives deterministic rotation, contouring the stationary distribution. In the case of constant isotropic noise, $\bm{D}= \epsilon \bm{I}$, \Rev{the drift can be written as the sum of a divergence-free curl flux ($\nabla \cdot \bm{J}_{ss}/P_{ss} = 0$) and the gradient of a scalar potential},
\begin{align}
    \bm{F} = \bm{J}_{ss}/P_{ss} + \nabla \log \epsilon P_{ss}.
\end{align}
As a result, the process is reversible if and only if the drift field is \textit{conservative} i.e. it is the gradient of a scalar potential \cite{Qian2002thermodynamics}. In the case of non-isotropic noise, we can perform a coordinate change $\mathbf{x} = \bm{D}^{-1/2}\mathbf{y}$ that yields a process with isotropic noise \cite{strang2020applications}.\\
\\
We also introduce the HHD of a flow on the edges of a simplicial complex \cite{Lim2020Hodge}. Given a graph $G = (V,E)$, where $V$ is a finite set of vertices and $E$ is the set of edges, higher-order cliques such as the set of triangles (3-cliques) are automatically defined, yielding a simplicial complex. Real-valued functions defined on the $k-$cliques of a simplicial complex are denoted \textit{flows} with the constraint that, for $k>1$, flows must be \textit{alternating} e.g. $X(i,j)=-X(j,i)$ is an alternating edge-flow (see Ref. \cite{Lim2020Hodge} for further details). Typically the HHD is formulated as a decomposition of an edge-flow into three orthogonal components, a gradient flow defined by a potential on the vertices, local circulations around triangles in the complex, and the remaining harmonic flow associated with the Hodge Laplacian \cite{Lim2020Hodge,Schaub2020RandomWalks}. Loosely speaking, the dimension of the harmonic space, the nullity of the Hodge Laplacian, is given by the number of \textit{holes} in the complex \cite{Lim2020Hodge}. However, by loosening the restriction that circulations must take place on triangles and allowing them to take place over longer loops, the harmonic contribution becomes zero \cite{strang2020applications}. In our approach, we will consider the \textit{Delauney triangulation} \cite{carlsson2009topological,loera2010triangulations} which automatically defines a contractible complex with no holes, thereby eliminating harmonic flow. Thus, by introducing discrete analogues of the curl, divergence and gradient operators (and their adjoints) \cite{grady2010discretecalculus}, we can decompose an edge-flow, $X$, as the sum of two orthogonal components,
\begin{align}
    X & = \curl^* \Phi +  \grad f, \label{eq: WHHD}
\end{align}
where $\Phi$ is an alternating function defined on the triangles, $f$ is a function defined on the vertices, with $\Phi$ and $f$ being analogous to a vector and scalar potential respectively (see App. \ref{app: HHD} for details) \cite{Lim2020Hodge,strang2020applications}. Therefore an edge-flow on a triangulation can be uniquely decomposed into a curling component and the gradient of a potential defined on the vertices, mirroring the decomposition of the stationary Langevin process, where, informally, the term $\grad f$ mirrors the gradient term in Eq. \ref{eq: helmholtzsde}, thus $\exp(f/\epsilon)$ provides a natural estimate of the stationary distribution. Similarly, $\curl^* \Phi$ captures the irreversible probability currents of Eq. \ref{eq: helmholtzsdeirrev}.\\
\\
Finally, we consider the HHD of a discrete-state Markov process which can be formulated as a flow on the edges of graph. We begin with a master equation of the form,
\begin{align}
    \frac{d\bm{p}}{dt} &= \bm{Q}\bm{p}\label{eq: master}
\end{align}
where $\bm{p} = (p_1,...,p_M)$ represents the probability of a random walker being at each of $M$ vertices and $\bm{Q}=(q_{ij})$ is the $Q$-matrix (also referred to as the generator or Laplacian matrix) of the continuous-time Markov chain defined by a random walk restricted to the edges of the graph. Off diagonal entries $q_{ji}>0$ of the $Q-$matrix represent \textit{transition intensities} between vertices $i$ and $j$ whilst diagonal entries are defined to be $q_{ii} = -\sum_j q_{ji}$, thus each column sums to 0 \cite{suhov2008markov}. Furthermore, we assume every transition is reciprocated i.e. if $q_{ij}>0 $ then $q_{ji}>0$. Next, we define the edge-flow to be,
\begin{align}
    X_{ij} = \log \sqrt{\frac{q_{ij}}{q_{ji}}},
\end{align}
which is naturally alternating i.e. $X_{ij}=-X_{ji}$, and can be motivated by both mathematical and thermodynamic arguments (see Chapter 6 of Ref. \cite{strang2020applications}). In particular, it is analogous to the excess work required to move from state $i$ to $j$ \cite{strang2020applications, Qian2001functional}. Furthermore, this definition of the edge-flow is analogous to the drift field of the Langevin process, because the Markov process is reversible and obeys detailed balance if and only if $X$ is a conservative flow i.e. it is the gradient of a scalar potential defined on the vertices (see Chapter 6.3 of Ref. \cite{strang2020applications} for a proof). Therefore the HHD can yield insight into the reversible and irreversible dynamics of a discrete-state, continuous-time Markov process.
\subsection{\label{sec: ReconDecomp} Approximating force fields as flows on graphs}
\begin{figure*}
    \centering
    \includegraphics[width = \textwidth]{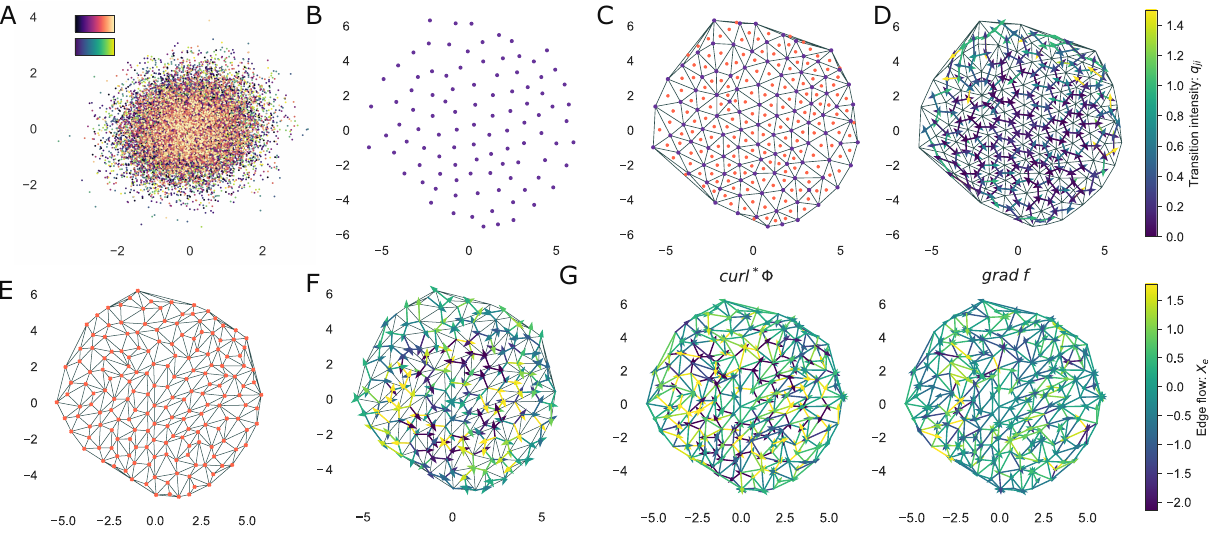}
    \caption{\textbf{Reconstruction and decomposition of drift fields as flows on graphs.} \textbf{A.} We begin with an ensemble of stochastic trajectories from the same system as a point-cloud on phase-space, \Rev{illustrated here with two trajectories}. \textbf{B.} We sub-sample the points using furthest point sampling to obtain a uniform density representation of phase-space. \textbf{C.} We discretise phase-space using a triangulation. Red points denote the center of each bin. \textbf{D.} We use maximum-likelihood estimation to reconstruct the transition intensities of a discrete-state Markov process with transitions between adjacent bins. \textbf{E.} Triangulating the midpoints of the bins, we obtain a simplicial complex representation of phase-space with transitions occurring over edges. \textbf{F.} Following the analysis in Sec. \ref{sec: HH}, we define the edge-flow by taking the logarithm of the square-root of the ratio of the transition intensities between connected bins. \textbf{G.} We perform the discrete HHD to obtain the curl and gradient components of the flow on the complex. \Rev{The edge-flow colourbar refers to panels F and G.}}
    \label{fig: Pipeline}
\end{figure*}
\Rev{Our approach attempts to approximate an unknown, underlying continuous Langevin process with a discrete Markov process that is reconstructed from an ensemble of stochastic trajectories. Representing the Markov process as a flow on a triangulation, we can leverage the correspondence between the HHD of the continuous and discrete-state processes to perform the HHD in discrete-space thus decomposing the process into its reversible and irreversible parts. This will allows us to approximate the curling and gradient components of the continuous drift-field with a discrete counterpart that can be calculated in a computationally simple manner. This is due to the relative ease with which the HHD of an edge-flow can be computed through its formulation as a least-squares optimisation problem (see App. \ref{app: HHD} and Ref. \cite{Lim2020Hodge})}.\\\\
Our framework, outlined in Fig. \ref{fig: Pipeline}, begins with an ensemble of trajectories assumed to be from the same stationary Langevin process. In this first case we assume that the noise is known to be isotropic with given $\epsilon$. Subsequently, we will treat the case of non-isotropic noise. Trajectories live in phase-space but are often clustered onto a particular, potentially lower dimensional, manifold \cite{fefferman2016manifold}. In order to discretise phase-space optimally given our observed trajectories, we consider them to be a point-cloud and perform furthest-point sampling \cite{moenning2003furthest} which yields a uniformly populated representation of phase-space that is optimal for manifold learning \cite{Gosztolai2023interpretable}. Tessellating the sub-sampled points with the Delauney triangulation, we obtain an irregular, triangular `grid' of $M$ bins (for an introduction to Delauney triangulations see Ref. \cite{loera2010triangulations}). We aim to approximate the Langevin process with the master equation defined in Eq. \ref{eq: master}. The discrete state of the process is the triangle/bin in which the trajectory is in at each time-point. After converting trajectories to observations of the discrete-state process, we use the maximum-likelihood estimator to estimate the transition intensities \cite{billingsley1961markov},
\begin{align}
    \hat{q}_{ji}& = \frac{N_{i \rightarrow j}}{T_i},
\end{align}
where $N_{i \rightarrow j}$ is the observed number of transitions between adjacent triangular bins $i$ and $j$, $T_i$ is the total `holding' time spent in each bin and bins are considered to be adjacent if they share an edge. Additionally, we set $\hat{q}_{ii} = -\sum_{j}\hat{q}_{ji}$. Given sufficient time-resolution, transitions should only occur between adjacent bins, thus transitions between non-adjacent bins that arise due to insufficient sampling frequency are not included in the estimation. Moreover, we include a pseudo-count, by initialising all $N_{i \rightarrow j}$ with a single unobserved transition, which eliminates singularities due to limited observations and which corresponds to a uniform Bayesian prior for the intensities \cite{manning2008laplacesmoothing}.\\\\
Next, we obtain the dual triangulation of the original grid by using the midpoints of the bins as vertices and performing the Delauney triangulation again. As a result, each vertex corresponds to a discrete state/bin in the Markov process and adjacent bins/reachable states are now connected by edges. \Rev{The flow on the edges of this second triangulation is defined to be $X_{ij} = \log \sqrt{q_{ij}/q_{ji}}$ where we assume the natural orientation $(i,j)$ if $i<j$. This flow is the discrete counter-part to the drift field as its curling component represents irreversible flow and its gradient component represents reversible flow. More specifically, the Markov process obeys detailed balance iff this flow is conservative \cite{strang2020applications}.}
\subsubsection{Non-isotropic diffusion}
\label{subsec: nonisotropic}
In the case of a process with non-isotropic diffusion, we must first perform the coordinate change $\mathbf{x} = \bm{D}^{-1/2}\mathbf{y}$ to obtain a process, $\mathbf{y}(t)$, with isotropic diffusion. In order to infer the diffusion matrix from the stochastic trajectories, we utilise the original tessellation and calculate the maximum-likelihood estimator \cite{Friedrich2011complexity},
\begin{align}
\label{eq: diffusion}
    \hat{\bm{D}}(\Delta_{\bar{\mathbf{x}}}) &=\frac{1}{N(\Delta_{\bar{\mathbf{x}}})}\sum_{i,r\;:\; \mathbf{x}_{t_i,r}\in \Delta_{\bar{\mathbf{x}}}}\frac{[\mathbf{x}_{t_{i+1},r} - \mathbf{x}_{t_{i},r}]\otimes [\mathbf{x}_{t_{i+1},r} - \mathbf{x}_{t_{i},r}]}{2\tau},
\end{align}
where $N(\Delta_{\bar{\mathbf{x}}})$ represents the number of points in each bin $\Delta_{\bar{\mathbf{x}}}$ identified by its mid-point $\bar{\mathbf{x}}$, $\{\mathbf{x}_{t_1,r},...,\mathbf{x}_{t_T,r}\}_{r=1}^{R}$ is the ensemble of trajectories, $\otimes$ is the outer product and $\tau$ is the time-step between observations. As the diffusion is assumed to be spatially homogeneous, given by a constant matrix, we take the mean value of the reconstructed matrices, $\bm{D}= \sum_{\Delta}\bm{D}(\Delta)/M$, which we can then use to transform the coordinates of the stochastic trajectories and apply the original method. In some physical systems, such as the dynamics of soft-matter in a medium, spatial-dependence in the diffusion tensor is a central feature \cite{Lau2007statedependent}. In the real-world systems considered here, the diffusion represents measurement error which we consider to be state-independent and temporally uncorrelated \cite{diterlizzi2024variancesum,qu2014heart}. In App. \ref{app: data isotropic}, we briefly investigate the validity of this assumption by assessing if there is spatially-dependent diffusion in the real-world data considered in this study.
\section{\label{sec: Res}Results}
\subsection{\label{subsec: numerics}Numerical experiments}
\begin{figure*}
    \centering
    \includegraphics[width = \textwidth]{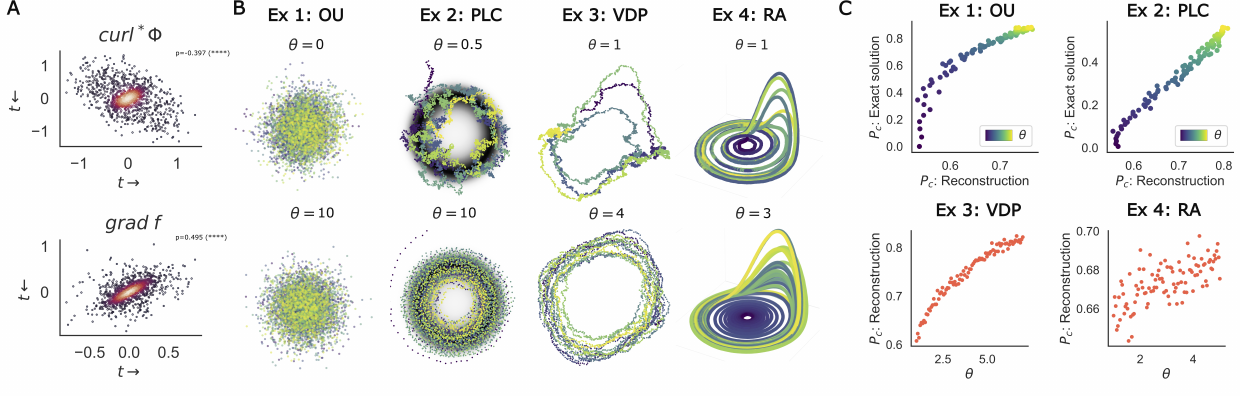}
    \caption{\textbf{Numerical experiments.} \textbf{A.} We begin with an ensemble of stochastic trajectories from a nonequilibrium OU process and apply our approach to both the original and time-reversed trajectories. As the curl part is odd under time-reversal, we see a significant negative correlation (-0.397, $p<0.0001$) between the curl flow along each edge calculated with the original and time-reversed trajectories. Conversely, the gradient flow is even under time-reversal, thus the gradient flows are positively correlated $(0.495, p<0.0001)$. \textbf{B.} Example trajectories from each of the four example processes with a low and high level of the rotation parameter $\theta$. \textbf{C.} For the solvable examples (OU, PLC) we increase $\theta$, calculate the proportion of flow in the curl space and compare to the exact HHD. We show a strong qualitative agreement with the reconstructed proportion of curl flow increasing with $\theta$. The disparity between the overall proportions arises from finite data and the approximation of the continuous space with discrete bins. For the unsolvable models (VDP, RA), we cannot compare to the exact solution, however $\theta$ acts as a proxy parameter for the level of rotation in the drift field. For both systems, the proportion of the reconstructed flow that is in the curl space increases with $\theta$, indicating that the edge-flow and its HHD capture the difference between irreversible and reversible currents.}
    \label{fig: Numerics}
\end{figure*}
In order to validate that our approach does, in fact, capture curl and gradient flows from stochastic trajectories, we first apply our analysis to simulations of four canonical dynamical systems with additive isotropic diffusion. We begin with two solvable models where the stationary probability distribution can be written in closed form and thus the HHD can be performed exactly. Specifically, we consider the multivariate Ornstein-Uhlenbeck (OU) process \cite{Godreche2018OU}, a linear nonequilibrium process, and the stochastic planar limit cycle (PLC) \cite{Yuan2017decomposition}, a nonlinear, nonequilibrium process. Next, we consider two nonlinear, unsolvable models, the stochastic van der Pol oscillator (VDP) \cite{strogatz2008nonlinear}, a system characterised by an asymmetric limit-cycle, and the stochastic Rössler attractor (RA) \cite{Rossler1976attractor}, a 3-dimensional system with a chaotic attractor. Panel B in Fig. \ref{fig: Numerics} shows example trajectories from each of the four processes with two levels of rotation in the drift field. As detailed in App. \ref{App: Processes}, the systems are modified to be parameterised by $\theta$, which controls the level of rotation in the drift field, and $\epsilon$ which controls the rate of diffusion.\\\\
The irreversible/reversible components of the drift are odd/even under time-reversal respectively \cite{DaCosta_2023}. As a first assessment of the method, we construct and decompose edge-flows from trajectories and time-reversed trajectories of the nonequilibrium OU. Panel A of Fig. \ref{fig: Numerics} shows the correlation of the curl (top) and gradient (bottom) flows along each edge between the original and time-reversed trajectories, noting that both utilise the same triangulations of phase-space. The curl flows are negatively correlated (-0.397, $p<0.0001$) whilst the gradient flows are positively correlated (0.495, $p<0.0001$), indicating that the components of the HHD of the edge-flow qualitatively capture the odd/even behaviour of the components of the underlying Langevin process under time-reversal. Next, as a measure of the prevalence of irreversible currents in the process, we calculate the proportion of the total flow that is in the curl space i.e. $P_c =  ||\bm{X}_c||/[ ||\bm{X}_c||+||\bm{X}_g||]$ where $\bm{X}_c, \bm{X}_g$ represent curl and gradient components respectively. Panel C in Fig. \ref{fig: Numerics}, shows the proportion of the flow in the curl space for increasing values of $\theta$ for each process. For the OU and PLC, we compare $P_c$ for the exact HHD of the drift field, calculated using a finite bounded domain, and $P_c$ in the reconstructed edge-flow. As $\theta$ is increased, the proportion of the flow that is irreversible/curl increases in the exact HHD, an effect that is captured with strong agreement by the discrete reconstruction. For the VDP and RA processes, the exact HHD cannot be performed, thus we use $\theta$ as a proxy measure for the level of curl in the field, showing that $P_c$ increases with $\theta$ in the reconstructed flow. As a result, the discrete reconstruction and decomposition is able to capture the difference between reversible gradient flows and irreversible circulations which can be used to identify the prevalence of irreversible currents. Whilst the general increase in $P_c$ agrees between the exact HHD and the discrete reconstruction, we note that the reconstructed proportion does not perfectly mirror the exact solution. This effect could stem from a combination of the discrete approximation of phase-space, the finite length of the trajectories and, potentially, from an inconsistency between the continuous and discrete forms of the HHD. A brief analysis of the effect of variation of the method parameters is presented in App. \ref{app: variation}. However, the method is able to differentiate between comparable systems with different proportions of irreversible curl flow, which is necessary for the following real-world applications. Details on the processes, simulation techniques, simulation parameters and variation of the parameters are given in App. \ref{App: Processes}, \ref{App: Sampling}, \ref{app: numerics}, \ref{app: variation} and \ref{app: OU}.
\subsection{\label{subsec: Real-world examples}Real-world examples}
\begin{figure*}
    \centering
    \includegraphics[width = \textwidth]{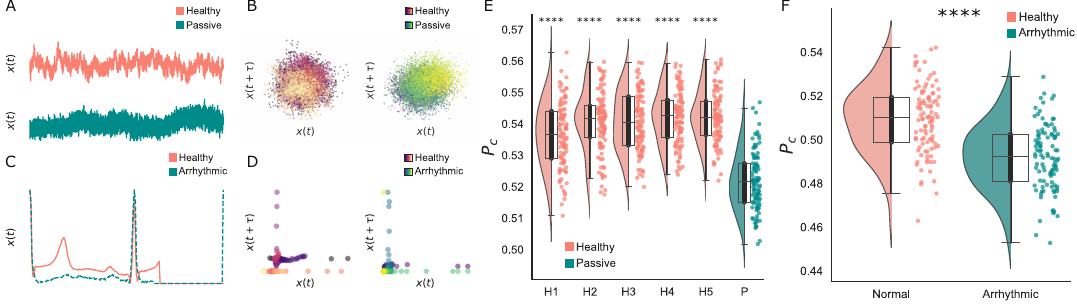}
    \caption{\textbf{Analysis of real-world data.} \textbf{A.} An example position trace for the outer membrane of a healthy (red) and passive (blue) RBC. \textbf{B.} An example trajectory in phase-space reconstructed using time-delay coordinates for both a healthy and passive RBC. \textbf{C.} An example ECG trace for a healthy participant (red) and one suffering myocardial infarction (blue). \textbf{D.} An example trajectory in phase-space reconstructed using time-delay coordinated for both a healthy and arrhythmic heartbeat. \textbf{E.} The proportion of flow in the curl space, $P_c$, for five healthy RBC and one passive RBC using 100 bootstrapped ensembles. We find that the prevalence of irreversible currents is significantly higher (****, $p<0.0001$) in each healthy cell when compared to the passive one. \textbf{F.} We compare healthy and arrhythmic heartbeats and find that $P_c$ is significantly higher (****, $p<0.0001$) in healthy heartbeats compared to arrhythmic ones, using 100 bootstrapped ensembles.}
    \label{fig: Data}
\end{figure*}
Noisy time-series from real-world systems can be modelled as stochastic trajectories from Langevin processes. In this section we consider two experiments with recordings of biophysical processes, namely flickering RBCs \cite{diterlizzi2024variancesum,turlier2019unveiling} and healthy and arrhythmic heartbeats \cite{goldberger2000physionet}. In the case of the RBCs, the dissipation of energy created by irreversible currents has been found to be lower in ATP-depleted cells compared to healthy cells \cite{diterlizzi2024variancesum,rodriguezgarcia2015directcytoskeleton}. As a result, the flickering recorded from passive cells is closer to thermal noise, and thus has a lower level of irreversible curling. Panel A of Fig. \ref{fig: Data} shows an example trace of the displacement of a point on the outer membrane of both a healthy and passive cell. The dynamics of the inner-membrane remains an unobserved variable, thus we assume the position trace is a scalar measurement of a 2-dimensional system \cite{diterlizzi2024variancesum}. Nonequilibrium dynamics have also been closely linked to the healthy functioning of biological organs \cite{peng2019nonequilibrium}, in particular in the heart where decreased time-irreversibility is associated with ageing and disease \cite{costa2005heart,peng2019nonequilibrium}. We consider electrical signals from the heart recorded as a electrocardiogram (ECG) which captures the voltage of action potentials. We compare the dynamics of healthy heartbeats with arrhythmic ones from patients with a myocardial infarction \cite{goldberger2000physionet}. Panel C of Fig. \ref{fig: Data} shows a typical ECG trace of a healthy (red) and arrhythmic (blue) heartbeat. Whilst the ECG is a scalar measurement, the limit-cycle nature of human heartbeats indicates a second, unobserved variable \cite{qu2014heart}.\\\\
In order to reconstruct a 2-dimensional trajectory in phase-space from a scalar measurement, we use a 2-d time-delay embedding with a single lag, $\tau$, chosen for each experiment by the mean of the first zero-crossing of the auto-correlation, thus we obtain a 2-d trajectory $(x(t),x(t+\tau))$ \cite{kantz2003nonlinear} (see App. \ref{app: data timedelay} for details \Rev{and App.} \ref{app: data timedelayvariation} \Rev{for the effect of varying this parameter.}). Panels B and D of Fig. \ref{fig: Data} show examples of reconstructed trajectories for healthy and passive RBCs and healthy and arrhythmic heartbeats. Unlike the example processes in Sec. \ref{subsec: numerics}, we cannot assume the diffusion is isotropic. As described in Sec. \ref{subsec: nonisotropic}, we first sub-sample and triangulate the trajectory in the reconstructed phase-space and use Eq. \ref{eq: diffusion} to estimate the diffusion matrix. Next, we change coordinates, $\mathbf{y}= \bm{D}^{1/2}\mathbf{x}$, resample and re-triangulate to then perform the analysis assuming isotropic diffusion. In order to compare if the statistical structure of the trajectories in each condition is responsible for the variation in $P_c$, we perform two forms of significance testing. Firstly, we use ensemble bootstrapping \cite{battle2016brokendetailedbalance,Lynn2021brokendetailedbalance} with 100 bootstraps to estimate the uncertainty in $P_c$ in each case (see App. \ref{app: data bootstrap}). Additionally, we perform surrogate-testing with shuffled time-series where shuffling destroys the statistical structure whilst keeping the points in phase-space the same, thus generating a `null' series  \cite{Lynn2021brokendetailedbalance} (see App. \ref{app: data shuffling}). Furthermore, in App. \ref{app: data isotropic} we assess the validity of assuming spatially-homogeneous diffusion in these datasets. For further details on the datasets and the analysis see App. \ref{app: data}.\\\\
Using our approach, we find that the proportion of irreversible, curling flow, $P_c$, is significantly higher (****, $p<0.0001$) in the flickering of each of five healthy RBCs when compared directly to a passive one (independent $t-$test) as shown in panel E of Fig. \ref{fig: Data}. This result is in agreement with previous results measuring the entropy production rate, a measure of nonequilibrium, in the same data \cite{diterlizzi2024variancesum}. Next, we apply our approach to the healthy and arrhythmic heartbeats and find that $P_c$ is significantly higher (****, $p<0.0001$) in healthy heartbeats when compared to arrhythmic beats from patients with myocardial infarction, as shown in panel F of Fig. \ref{fig: Data}. Our results validate previous indications that time-irreversibility, nonequilibrium dynamics, and by extension, the prevalence of curling currents, are a key feature of healthy heart dynamics \cite{peng2019nonequilibrium, costa2005heart}.
\section{\label{sec: relatedwork} Related work and other approaches}
\Rev{The analysis of nonequilibrium dynamics in biological systems has received much recent interest \cite{fang2019nonequilibrium,Gnesotto2018brokendetailedbalance}, in particular in data-driven methods for analysing real-world trajectories \cite{battle2016brokendetailedbalance,Frishman2020stochasticforce,Friedrich2011complexity,bruckner2024learning}. One typical approach is to infer approximate drift and diffusion terms directly from data \cite{Friedrich2011complexity, elbeheiry2015inferencemap,garcia2018reconstruction,bruckner2020inferring,ferretti2020generallangevin}, however only one such method `Stochastic Force Inference' \cite{Frishman2020stochasticforce} allows for the natural HHD of the inferred drift as well as an estimate of the entropy production rate. Typically, methods aim to approximate a continuous field with a discrete approximation, but do not formulate a discrete-state stochastic process, as we do here. `Probability flux analysis' \cite{battle2016brokendetailedbalance} is another approach, designed for biological data, that attempts to approximate probability currents in a coarse-grained discrete state-space. This approach approximates the continuous process with this discrete-state counterpart, by inferring the probability flux between system states, yet it stops short of decomposing the process into its reversible and irreversible parts or measuring the overall level of nonequilibrium. More generally, many methods have been developed to analyse nonequilibrium flows by utilising the link between nonequilibrium dynamics and the irreversibility of stationary trajectories \cite{roldan2010dissipation,feng2008length} with applications in biology \cite{Roldán2021irreversibility} and neuroscience \cite{delafuente2023irreversibility}. Similarly to the quantity $P_c$, such methods can quantify the overall prevalence of nonequilibrium flows in stochastic trajectories, but yield no insight into the governing dynamics of the process, unlike the reconstruction of a discrete-state process.\\\\
The discrete HHD has found many applications in the analysis of hierarchical and cyclical structures in graphs and tournaments \cite{strang2022networkhhd,Lim2020Hodge} with some applications to dynamics and signal processing \cite{Schaub2020RandomWalks,Schaub2021SignalProcessing,Schaub2022SignalProcessing,Barbarossa2020topsig}. To this point, the discrete HHD has not been applied to Markov processes to gain insight into their thermodynamics with the exception of Ref. \cite{strang2020applications}. Nevertheless, connections can be made between the HHD and classical `cycle decomposition' approaches that are well-studied in the thermodynamics of graphs and Markov processes \cite{schnakenberg1976network,Jiang2004noneq,Altaner2012network}. Similarly, results in the convergence of the discrete and continuous HHD are limited to simple domains and dynamics \cite{strang2020applications}.\\\\
The approximation of vector fields with discrete counter-parts is a central challenge in numerical analysis, such as finite element calculus \cite{Arnold2010FiniteElement}. Such approaches typically involve discretisation of continuous domains for the numerical solution of deterministic equations and have little relation to our approach. One more closely related approach is `MARBLE' \cite{Gosztolai2023interpretable} which aims to learn the manifold structure of phase-space, which is approximated as a graph, and infer the vector field directly from recordings of neural signals - though the dynamics are assumed to be deterministic thus the vector field is not formulated as a discrete-state Markov process.} 
\section{\label{sec: Discussion} Discussion}
Noisy, fluctuating signals represent the prototypical data that can be recorded from a real-world system. Whilst Langevin processes and SDEs provide a mathematical framework to describe such systems, the development of data-driven approaches is key to gaining insight into stochastic trajectories with unknown drift and diffusive fields \cite{Friedrich2011complexity}. In this paper, we focus on the HHD of a Langevin process into its irreversible, curling component and its reversible gradient component. By utilising the analogous HHD of a flow on the edges of a graph, we develop a method that is able to reconstruct and then decompose a discrete-state Markov process directly from stochastic trajectories. By focusing on the proportion of the flow in the curl space, we validate that our method captures the relative prevalence of irreversible currents between similar systems with different levels of nonequilibrium. We assess the efficacy of our approach on sampled paths from both solvable and nonlinear processes and then apply it to real-world data from biophysical experiments. Our approach is able to capture the greater prevalence of nonequilibrium flows in the dynamics of healthy RBCs when compared to ATP-depleted, `passive' ones \cite{diterlizzi2024variancesum, rodriguezgarcia2015directcytoskeleton} as well as the greater prevalence of nonequilibrium flows in healthy heartbeats compared to arrhythmic heartbeats from patients with myocardial infarction \cite{peng2019nonequilibrium, costa2005heart, goldberger2000physionet}.\\\\
Our approach presents a new application of graph and topological signal processing to the analysis of stochastic trajectories and real-world, noisy time-series with a specific focus on biological and physiological recordings. In particular, our work builds on previous studies that highlight nonequilibrium and irreversible dynamics as central to the healthy functioning of mesoscopic \cite{battle2016brokendetailedbalance} and macroscopic \cite{costa2005heart,peng2019nonequilibrium,Lynn2021brokendetailedbalance,sanzperl2021nonequilibrium} dynamics in biological systems. Finally, our method draws a novel and important link between the discrete HHD and its continuous counterpart that remains to be explored at a more theoretical and analytical level in order to investigate under what conditions this convergence is exact and both decompositions consistent \cite{strang2020applications}. As a first application, we focused on the overall prevalence of irreversible circulations, but we note that this approach can be developed, particularly in conjunction with existing techniques for analysing the discrete HHD \cite{strang2022networkhhd,strang2024PTA}, to analyse the specific structure of the gradient and circular flows in phase-space to yield greater insight into underlying dynamics. Such an avenue hold great promise for the further development of graph and topological data analysis techniques for time-varying data recorded from complex systems.
\section*{Data and code availability}
The code supporting the findings in this paper is available at \url{https://github.com/rnartallo/decomposingflows}.\\
\\
The heartbeat data used in this study can be found at Ref. \cite{fazeli2018heartdata} as well as further details at \cite{goldberger2000physionet,kachuee2018heart,bousseljot1995cardiodat}. The red-blood cell data used in this study can be found at Ref. \cite{diterlizzi2024data} with further details at Ref. \cite{diterlizzi2024variancesum}.
\section*{Author contributions}
R.N.K developed the method with input from all other authors, performed research and drafted the manuscript. All authors discussed results and edited the manuscript.
\section*{Acknowledgements}
The authors would like to thank Pierre Ronceray, Michael Schaub, Vincent Grande and Alexis Arnaudon for their helpful comments and conversations.\\\\
R.N.K was supported by an EPSRC doctoral scholarship from grants EP/T517811/1 and EP/R513295/1. M.L.K was supported by the Centre for Eudaimonia and Human Flourishing funded by the Pettit and Carlsberg Foundations and by the Center for Music in the Brain (MIB), funded by the Danish National
Research Foundation (project number DNRF117).
R.L. was supported by the EPSRC grants EP/V013068/1 and EP/V03474X/1.

\bibliographystyle{unsrtnat}
\bibliography{reference}

\appendix
\section{Appendix}
\subsection{Details on the Helmholtz-Hodge decomposition}
\label{app: HHD}
For a vector field $\mathbf{F}\in C^1(V,\mathbb{R}^n)$ where $V\subseteq \mathbb{R}^n$ that decays sufficiently quickly, the HHD is a division of the vector field, that may not be unique, into two components,
\begin{align}
    \mathbf{F}(\mathbf{r}) &= -\nabla \Phi(\mathbf{r}) + \mathbf{R}(\mathbf{r}),
\end{align}
where $\Phi \in C^{2}(V,\mathbb{R})$ is a scalar potential, $\nabla$ is the gradient operator and $\mathbf{R}(\mathbf{r})$ is divergence-free i.e. $\nabla\cdot\mathbf{R} = 0$ \cite{GLOTZL2023127138}. In three dimensions, assuming the vector fields decay sufficiently quickly at infinity, the divergence-free force can be written as the curl of a vector potential $\mathbf{A}$, i.e. $\mathbf{R} = \nabla \times \mathbf{A}$.
\subsubsection{The HHD of a stationary Langevin process}
\Rev{As detailed in Eq. \ref{eq: langevin}, the Langevin process is given by,
\begin{align}
    \dot{\mathbf{x}}(t) & = \bm{F}(\mathbf{x}(t)) + \bm{\xi}(t),
\end{align}
where $\bm{F}$ is the drift field, whilst $\bm{\xi}$ are Gaussian fluctuations with spatially-homogeneous correlations given by $\langle \bm{\xi}(t),\bm{\xi}(t')\rangle = 2\bm{D}\delta(t-t')$ with $\bm{D}$ the constant, positive definite, noise-covariance matrix.
As shown in Eq. \ref{eq: FP} and Eq. \ref{eq: conservation}, the density dynamics can be written as a conservation of probability law,
\begin{align}
\frac{\partial P}{\partial t} + \nabla\cdot \bm{J} & = 0,
\end{align}
where $\bm{J}(\mathbf{x},t)$ is the probability flux defined as,
\begin{align}
    \bm{J} & = \bm{F}P - \nabla\cdot (\bm{D}P),
\end{align}
and $\nabla \cdot $ represents the divergence operator, which, when defined on a vector field $\bm{A}:\mathbb{R}^N \rightarrow \mathbb{R}^N$, returns a scalar quantity,
\begin{align}
    \nabla \cdot \bm{A} (\mathbf{x}) = \sum_i \partial \bm{A} / \partial x_i,
\end{align}
where $\mathbf{x} = (x_1,...,x_N)$. When defined on a tensor field, $\bm{D}:\mathbb{R}^{N} \rightarrow \mathbb{R}^{N}\times \mathbb{R}^{N}$, it returns a vector in $\mathbb{R}^N$ whose $i$-th entry is given by,
\begin{align}
    (\nabla \cdot \bm{D}(\mathbf{x}))_i & = \sum_{j}\frac{\partial D_{ij}(\mathbf{x})}{\partial x_j}.
\end{align}
At stationarity, i.e. $\partial P/\partial t = 0$, the density of the process remains constant whilst the trajectories still evolve in time. The probability flux at stationarity, $\bm{J}_{ss}$, is divergence-free, $\nabla \cdot \bm{J}_{ss} =0$. The drift at stationarity can be written as \cite{fang2019nonequilibrium},
\begin{align}\label{eq: HHD appendix}
    \bm{F} &=  \frac{\bm{J}_{ss}}{P_{ss}} + \bm{D}\cdot \nabla \log P_{ss} + \nabla\cdot \bm{D},
\end{align}
where $\nabla$ represents the gradient operator, which takes a scalar potential, $U: \mathbb{R}^N \rightarrow \mathbb{R}$ and returns a vector field,
\begin{align}
    \nabla U(\mathbf{x}) &= \sum_i\frac{\partial U}{\partial x_i} \mathbf{e}_i,
\end{align}
where $\mathbf{e}_i$ is the $i-$th unit vector i.e. $(\mathbf{e}_{i})_{j} = \delta_{ij}$. Eq. \ref{eq: HHD appendix} decomposes the drift field into the sum of a curl flux, $\bm{J}_{ss}/P_{ss}$, which is divergence-free, and the gradient of the scalar potential, $U = \log P_{ss}$, multiplied by the diffusion $\bm{D}$. Note that, whilst we assume spatially homogeneous diffusion, we include the term $\nabla \cdot \bm{D} (=0)$ for completeness. This term is also the correction factor between the Îto and Stratonovich formulations of spatially-dependent diffusions \cite{vanKampen1981itostratonovich}. As a result, the stationary distribution and diffusive field give us the unique HHD of the drift. Moreover, in the case of a isotropic diffusion, $\bm{D} = \epsilon \bm{I},$ the reversible dynamics are exactly the gradient of the scalar potential $\log \epsilon P_{ss}$, and the drift is given by,
\begin{align}
    \bm{F} = \bm{J}_{ss}/P_{ss} + \nabla \log \epsilon P_{ss}.
\end{align}
As a result, the detailed balance condition, $\bm{J}_{ss} \equiv \bm{0}$, which guarantees reversible dynamics and an equilibrium stationary distribution, is met if and only if the drift field is conservative, i.e. there exists a scalar potential $V$ such that $\nabla V = \bm{F}$. This condition can be used to define a discrete-analog of the drift for the discrete-state Markov process and discrete gradient operator.}
\subsubsection{The HHD of a edge-flow on a simplicial complex}
Given a simplicial complex with vertices, edges and triangles, $V,E,T$ respectively, we define the set of flows on each set to be $L^2(V), L^2_{\wedge}(E), L^2_{\wedge}(T)$ respectively, where $\wedge$ indicates that the flows are alternating. Next, we define inner-products on these spaces. We take unweighted inner products, $\langle f,g\rangle = \sum_{i}f(i)g(i)$, $\langle X,Y\rangle = \sum_{i<j}X(i,j)Y(i,j)$ and $\langle \Phi,\Theta\rangle = \sum_{i<j<k}\Phi(i,j,k)\Theta(i,j,k)$ for the spaces of vertices, edges and triangles respectively. Using these inner products, we define the \textit{gradient}, \textit{curl} and \textit{divergence} operators acting between these function spaces,
\begin{align}
    (\grad \;f)(i,j) &= f(j)-f(i),\\
    (\curl \;X)(i,j,k) &= X(i,j) + X(j,k) + X(k,i),\\
    (\dive \;X)(i) &= \sum_j X(i,j).
\end{align}
where $\grad:L^2_{\wedge}(V)\rightarrow L^2_{\wedge}(E)$,  $\curl:L^2_{\wedge}(E)\rightarrow L^2_{\wedge}(T)$ and $\dive:L^2_{\wedge}(E)\rightarrow L^2_{\wedge}(V)$ \cite{Lim2020Hodge}. From these definitions, we are able to define the curl adjoint, $\curl^*: L^2_{\wedge}(T)\rightarrow L^2_{\wedge}(E)$, which is given by,
\begin{align}
     (\curl^* \;\Phi)(i,j) &= \sum_k \Phi(i,j,k).
\end{align}
Thus, these operators define the following relationship between the function spaces,
\begin{align}
    L^2_{\wedge}(V) \underset{\text{div}}{\stackrel{\text{grad}}{\rightleftarrows}}   L^2_{\wedge}(E) \underset{\text{curl}^*}{\stackrel{\text{curl}}{\rightleftarrows}} L^2_{\wedge}(T).
\end{align}
Additionally, we can define the \textit{graph-Helmholtzian}, $\Delta_1: L^2_{\wedge}(E)\rightarrow L^2_{\wedge}(E)$, a discrete analogue of the vector Laplacian, defined by,
\begin{align}
\label{eq: helmholtzian}
    \Delta_1 &= -\grad\; \dive + \curl^*\;\curl.
\end{align}
The HHD associated with this operator states that the space of edge-flows admits a decomposition into orthogonal subspaces,
\begin{align}
    L^2_{\wedge}(E) = \imag(\curl^*) \bigoplus \imag(\grad) \bigoplus \ker(\Delta_1).
\end{align}
As a result, the HHD states that any edge-flow $X$ can be decomposed as follows,
\begin{align}
    X & = \curl^* \Phi + \grad f + X_H,
\end{align}
where $\curl^* \Phi$ is a divergence-free flow which is the curl-adjoint of a flow, $\Phi$, defined on the triangles, $\grad f$ is the curl-free gradient of a flow on the vertices and $X_H$ is a harmonic flow on the edges i.e. $(\Delta_1(X_H) = \mathbf{0})$. A gradient flow is induced by assigning a potential to each of the nodes and then defining the edge-flow according to the difference of the potentials along each edge, thus gradient flows have zero net sum along any closed path in the graph. The space of non-gradient flows contains cyclical flows that are either curling, representing local circulations around triangles, or harmonic, representing global circulations around \textit{holes} in the complex \cite{Schaub2022SignalProcessing}. As there are no holes in the Delauney triangulation, the harmonic flow is zero \cite{Lim2020Hodge}, thus we see the HHD that we will aim to calculate is,
\begin{align}
    X & = \curl^* \Phi + \grad f,
\end{align}
which mirrors the decomposition of the stationary Langevin process with isotropic diffusion.
\\\\
The HHD of a edge-flow can be computed efficiently as a pair of sparse least-squares problems, one for the gradient component and one for the curling component \cite{Lim2020Hodge}. As we have no harmonic flow, we need only solve one problem and obtain the other component as the residual of the edge-flow and first component. We will solve for the gradient component. We first note that, the space of functions on vertices, edges and triangles are each finite-dimensional and permit the following isomorphisms $L^2(V)\cong \mathbb{R}^{|V|}$, $L^2_{\wedge}(E)\cong \mathbb{R}^{|E|}$, and $L^2_{\wedge}(T)\cong \mathbb{R}^{|T|}$,
where $|V|,|E|,|T|,$ are the number of vertices, edges and triangles respectively. As a result, flows can be uniquely expressed as a vector whilst the discrete operators that project between these spaces can be expressed as sparse matrices. Utilising that the decomposition is orthogonal, this allows us to compute the decomposition by solving the sparse least-squares problem with an algorithm such as LSQR \cite{Paige1982LSQR},
\begin{align}
    f & = \argmin_{g\in \mathbb{R}^{|V|}}||\grad \; g - X||^2
\end{align}
We can recover the curl part of the flow by taking,
\begin{align}
    \curl^* \Phi = X- \grad\; f.
\end{align}
\subsection{\label{App: Processes}Example stochastic processes}
\subsubsection*{Example 1. The multivariate Ornstein-Uhlenbeck process}
Our first example process is a linear Langevin equation,
\begin{align}
    \frac{d\mathbf{x}}{dt}& = -\left [ \begin{pmatrix}
2 & 0\\ 
0 & 2
\end{pmatrix} + \begin{pmatrix}
0 & -\theta\\ 
\theta & 0
\end{pmatrix}\right ]\mathbf{x} + \bm{\xi},
\end{align}
with isotropic diffusion given by,
\begin{align}
     \bm{D}&=\epsilon \bm{I}.
\end{align}
The parameter $\theta$ controls the level of rotation in the drift with increases in $\theta$ resulting in increased rotation and $\theta = 0$ corresponding to a reversible process with only gradient flow. The diffusion parameter $\epsilon$ controls the intensity of the noise. The stationary distribution of this process is a zero-mean multivariate Gaussian, thus it permits an exact HHD (for a general treatment of the OU process see App. \ref{app: OU} and Ref. \cite{DaCosta_2023,Godreche2018OU}). The system also permits an exact simulation from its transition probability density (see App. \ref{App: Sampling}) \cite{DaCosta_2023}.
\subsubsection*{Example 2. The stochastic planar limit cycle}
Next, we consider the nonlinear process,
\begin{align}
    \dot{x}&= -\theta y + x(1-x^2-y^2)+ \xi_x,\\
\dot{y}&= \theta x + y(1-x^2-y^2)+ \xi_y,
\end{align}
with isotropic diffusion given by $\langle \xi_i(t),\xi_j(t') \rangle = 2\epsilon \delta_{ij}\delta(t-t')$. In the absence of diffusion, this system converges to a stable circular limit-cycle \cite{Yuan2017decomposition}. Additionally, it admits a `Mexican-hat' potential of the form $\Phi(x,y) = (x^2 + y^2)(x^2+y^2-2)/4\epsilon$ which is associated with the Boltzmann-Gibbs stationary distribution,
\begin{align}
    P_{ss}(x,y)&=\frac{1}{Z_\epsilon}\exp\left(-\frac{(x^2 + y^2)(x^2+y^2-2)}{4\epsilon}\right),
\end{align}
where the partition function is $Z_{\epsilon}=e^{1/4\epsilon}\sqrt{\epsilon}\pi^{3/2}(1+\erf[1/(2\sqrt{\epsilon})])$ \cite{Yuan2017decomposition}. Given the stationary distribution, we decompose the drift into,
\begin{align}
\bm{F}_{\text{rev}}& = \begin{pmatrix}
x(1-x^2-y^2)\\ 
y(1-x^2-y^2)
\end{pmatrix},\;\;\bm{F}_{\text{irrev}}=\begin{pmatrix}
0 & -\theta\\ 
\theta & 0
\end{pmatrix}\begin{pmatrix}
x\\ 
y
\end{pmatrix}.
\end{align}
The parameter $\theta$ controls the degree to which the trajectories rotate and thus contour the stationary distribution. In order to sample from this model we avoid traditional finite-difference schemes, such as the Euler-Maruyama scheme, which do not preserve the ergodic, oscillatory or nonequilibrium dynamics of stochastic systems \cite{mattingly2002ergodicity,buckwar2022FHN,DaCosta_2023}. Instead, we employ geometric splitting integration (see \ref{App: Sampling}) \cite{hairer2006geometric} that preserves the structure of the stochastic oscillatory dynamics \cite{buckwar2022FHN}.
\subsubsection*{Example 3. The stochastic modified van der Pol oscillator}
We consider a modified version of the stochastic VDP limit-cycle oscillator \cite{strogatz2008nonlinear},
\begin{align}
    \dot{x}&= \theta y +\xi_x,\\
\dot{y}&= -\theta x + \mu y(1-x^2) + \xi_y,
\end{align}
with diffusion given by $\langle \xi_i(t),\xi_j(t') \rangle = 2\epsilon \delta_{ij}\delta(t-t')$. In the absense of noise, this system converges to an asymmetric limit-cycle. Whilst we do not have access to the stationary distribution and the exact HHD in this case, we can use that increasing $\theta$ creates a stronger rotational component to the drift field as a approximate control for the curling part of the drift. Setting $\theta=1$ corresponds to original VDP oscillator.
\subsubsection*{Example 4. The stochastic modified Rössler attractor}
Finally, we consider a modified version of the Rössler system \cite{Rossler1976attractor} given by the 3-dimensional system,
\begin{align}
    \dot{x}&= -\theta y - z +\xi_x,\\
\dot{y}&= \theta x +ay + \xi_y,\\
\dot{z}&= b + z(x-c) + \xi_z,
\end{align}
where the diffusion is given by $\langle \xi_i(t),\xi_j(t') \rangle = 2\epsilon \delta_{ij}\delta(t-t')$. With the parameters $a = 0.2$, $b = 0.2$, $c=5.7$,  $\theta=1$ and $\epsilon = 1$, the system exhibits attracting chaotic dynamics. The parameter $\theta$ is a modification that controls the level of rotation in the $xy-$plane. We sample from this process using the mean-square convergent \textit{tamed Euler-Maruyama} scheme (see \ref{App: Sampling}) \cite{hutzenthaler2012tamed}.
\subsection{Sampling from Langevin processes}
\label{App: Sampling}
In Sec. \ref{subsec: numerics}, we validate our approach with trajectories sampled from a range of Langevin processes. In order for such simulations to validate the theory, we take care to use simulation techniques that preserve the underlying structure of the dynamics and stationary distribution.
\subsubsection{Exact simulation of the Ornstein-Uhlenbeck}
Due to its linearity, the OU process permits an exact simulation from its stationary distribution \cite{DaCosta_2023}. Given a Langevin process,
\begin{align}
    \frac{d\bm{x}}{dt}&=-\bm{B}\mathbf{x}(t)+ \xi(t),
\end{align}
with $\langle\bm{\xi}(t),\bm{\xi}(t')\rangle = 2\bm{D}\delta(t-t')$, the solution is given by,
\begin{align}
    \mathbf{x}(t) = e^{-\bm{B}t}\mathbf{x}(0) + \int_{0}^{t}e^{-\bm{B}(t-s)}\Sigma \;d\bm{W}(s),
\end{align}
where $\Sigma\Sigma^{\top} = 2\bm{D}$. Therefore, the transition kernel is given by,
\begin{align}
    \mathbf{x}(t) \sim \mathcal{N}( e^{-\bm{B}t}\mathbf{x}(0),\bm{C}(t)),
\end{align}
where,
\begin{align}
    \bm{C}(t) &= \int_0^t2e^{-\bm{B}(t-s)}\bm{D}(e^{-\bm{B}(t-s)})^{\top}\;ds,
\end{align}
thus the process can be simulated exactly.
\subsubsection{Splitting integrators for oscillatory processes}
In general, nonlinear processes do not admit exact simulation and we must employ numerical methods to simulate trajectories. However, standard techniques such as the Euler-Maruyama scheme do not preserve the underlying dynamics of stiff oscillatory systems \cite{buckwar2022FHN} and are not mean-square convergent \cite{mattingly2002ergodicity}, thus are unsuitable. We employ a geometric splitting integrator for oscillatory SDEs \cite{hairer2006geometric,buckwar2022FHN} which preserves the underlying dynamics, in order to simulate from both the PLC and the VDP. Given a nonlinear Langevin process,
\begin{align}
    \frac{d\bm{x}}{dt}&=F(\mathbf{x}(t))+ \xi(t),
\end{align}
with $\langle\bm{\xi}(t),\bm{\xi}(t')\rangle = 2\bm{D}\delta(t-t')$, we `split' the drift term into a nonlinear, deterministic part and a linear stochastic part \cite{buckwar2022FHN},
\begin{align}
    \mathbf{x}^{[1]}(t)&=\bm{A}\mathbf{x}(t) +\xi(t),\\
    \mathbf{x}^{[2]}(t)&=\bm{G}(\mathbf{x}(t)),
\end{align}
where $\bm{F}(\mathbf{x})=\bm{A}\mathbf{x} + \bm{G}(\mathbf{x})$. The linear part, $\mathbf{x}^{[1]}$, defines an OU process which can be simulated exactly. If the splitting is chosen such that the ordinary differential equation for $\mathbf{x}^{[2]}$ can be solved explicitly, then the solution of the full process can be approximated with the Strang approximation \cite{hairer2006geometric},
\begin{align}
    \mathbf{x}(t+h) \approx \varphi_1^{[h/2]}\circ \varphi_2^{[h]} \circ \varphi_1^{[h/2]} (\mathbf{x}(t)),
\end{align}
where $\varphi_i^{[h]}$ is a simulation from the exact $h$-time transition kernel of $\mathbf{x}^{[i]}$.\\\\
For the PLC, we use the decomposition,
\begin{align}
    \mathbf{x}^{[1]}(t)&=\begin{pmatrix}
 1&-\theta \\ 
\theta & 1
\end{pmatrix}\mathbf{x}(t) +\xi(t),\\
    \mathbf{x}^{[2]}(t)&=\begin{pmatrix}
-(x^2+y^2)x\\ 
-(x^2+y^2)y
\end{pmatrix},
\end{align}
where the nonlinear part can be solved easily in polar coordinates.\\\\
For the VDP, we use the decomposition
,
\begin{align}
    \mathbf{x}^{[1]}(t)&=\begin{pmatrix}
 0&\theta \\ 
-\theta & \mu
\end{pmatrix}\mathbf{x}(t) +\xi(t),\\
    \mathbf{x}^{[2]}(t)&=\begin{pmatrix}
0\\ 
-\mu x^2y
\end{pmatrix},
\end{align}
where the nonlinear part can be solved trivially due to the constancy of the first component $x$. 
\subsubsection{Tamed Euler-Maruyama for the Rössler attractor}
Due to the difficulties in decomposing the Rössler system, we instead opt for the mean-square convergent, tamed Euler-Maruyama scheme \cite{hutzenthaler2012tamed},
\begin{align}
    \mathbf{x}(t+h) \approx \mathbf{x}(t) + h \frac{\bm{F}(\mathbf{x}(t)}{1+||\bm{F}(\mathbf{x}(t)||} + \Sigma \bm{\nu}_h,
\end{align}
where $\Sigma\Sigma^{\top}=2\bm{D}$ and $\bm{\nu}_h$ is a Gaussian vector with independent components each with zero-mean and variance $h$.
\subsection{Parameters for numerical simulations and data analysis}
\label{app: numerics}
For the numerical experiments presented in panel C of Fig. \ref{fig: Numerics}, we utilised 500 trajectories of length 10,000 for each process. For the OU process, we utilised $h=0.01, \epsilon = 0.5$ with 250 sub-sampled points. We swept $\theta$ between 0 and 15 with 101 equispaced values. For the PLC process, we utilised $h=0.01, \epsilon = 0.25$ with 250 sub-sampled points. We swept $\theta$ between 0 and 10 with 100 equispaced values. For the VDP process, we utilised $h=0.01, \epsilon = 0.25$ with 250 sub-sampled points. We swept $\theta$ between 1 and 7 with 90 equispaced values. For the RA process, we utilised $h=0.01, \epsilon  =0.075$ with 250 sub-sampled points. We swept $\theta$ between 1 and 4 with 100 equispaced values. For the analysis of both RBCs and heartbeats, we used 250 sub-sampled points. The diffusion estimator used an arbitrary time-step of $h=0.01$.
\subsection{Variation of method parameters}
\label{app: variation}
\begin{figure*}
    \centering
    \includegraphics[width = \textwidth]{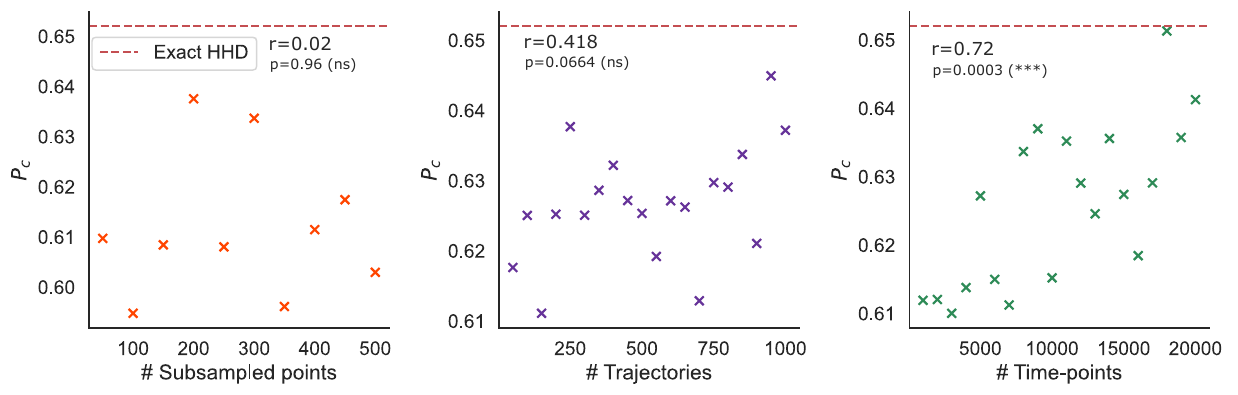}
    \caption{\textbf{Variation of method parameters.} Here we show the effect of varying the number of sub-sampled points used to construct the grid and graph, the number of trajectories in an ensemble and the number of time-points in a trajectory, using the OU process with $\theta = 4$ as an example. The estimate from the exact decomposition at $\theta = 4$ is 0.652 (3.s.f), \Rev{as shown by the red dashed line}.}
    \label{fig: sensitivity}
\end{figure*}
The method we propose depends on a number of key parameters such as the number of sub-sampled points used to construct the triangulations, the number of trajectories in an ensemble and the number of time-points in a trajectory. In Fig. \ref{fig: sensitivity} we consider the OU process fixed at a value of $\theta=4$ and vary, independently, the number of sub-sampled points, the number of trajectories and the length of the trajectories. The relationship between these parameters is complex, unlike in typical numerical schemes. For example, simply increasing the number of sub-sampled points, thereby refining the grid, will result in fewer points per bin/triangle and a worse estimate of the transition intensity or diffusion matrix. On the other hand, using fewer points will result in coarse-grid that does not reflect state-space correctly. It is due to this added complexity, that we refrain from attempting a proof of convergence that would require taking both the infinite data and the infinite point limit. In Fig. \ref{fig: sensitivity}, we see that there is a fluctuation of $P_c$ with the the number of sub-sampled points, but no correlation. In the case of the number of trajectories or the length of the trajectories, we would expect increasing this parameter to result in an improved estimate of $P_c$. As shown in Fig. \ref{fig: sensitivity}, the value of $P_c$ appears to increase with both the number of time-points per trajectory ($p=0.003$, 3.s.f) and the number of trajectories ($p=0.0664$, 3.s.f) which implies that they are approaching the true value which can estimated to be $P_c =0.652$ (3.s.f) from the exact decomposition at $\theta = 4$.
\subsection{Details on real-world data}
\label{app: data}
\subsubsection{Time-delay embedding for phase-space reconstruction}
\label{app: data timedelay}
Phase-space coordinates can be reconstructed from scalar measurements via a time-delay embedding \cite{kantz2003nonlinear}. For both the RBC and heartbeat analysis, we have a univariate time-series recording a single scalar measurement from the system. In the case of the RBC, this is the displacement of a single point chosen on the outer membrane. For the heartbeat, this is an electrocardiogram (ECG). However, as mentioned in Sec. \ref{sec: Res}, both systems are truly two dimensional as the RBC has an unmeasured inner membrane displacement \cite{diterlizzi2024variancesum} whilst heartbeat dynamics are known to obey two dimensional limit cycle dynamics \cite{qu2014heart}. Assuming we have access to coordinate $x(t)$ with unseen coordinate $y(t)$, we approximate the phase-space $(x(t),y(t))$ with $(x(t),x(t+\tau))$ where $\tau$ is known as the \textit{lag} or \textit{delay}. In order to select $\tau$, we calculate the first lag at which the auto-correlation crosses 0 for each trajectory. We take the mean value over all trajectories in the dataset and fix $\tau$ for the remainder of the experiment. For the RBC experiments we obtain $\tau = 998, (T=10,000)$, whilst for the heartbeats we obtain $\tau = 45, (T=188)$.
\subsubsection{Spatially-dependent diffusion in real-world data}
\label{app: data isotropic}
\begin{figure*}
    \centering
    \includegraphics[width = \textwidth]{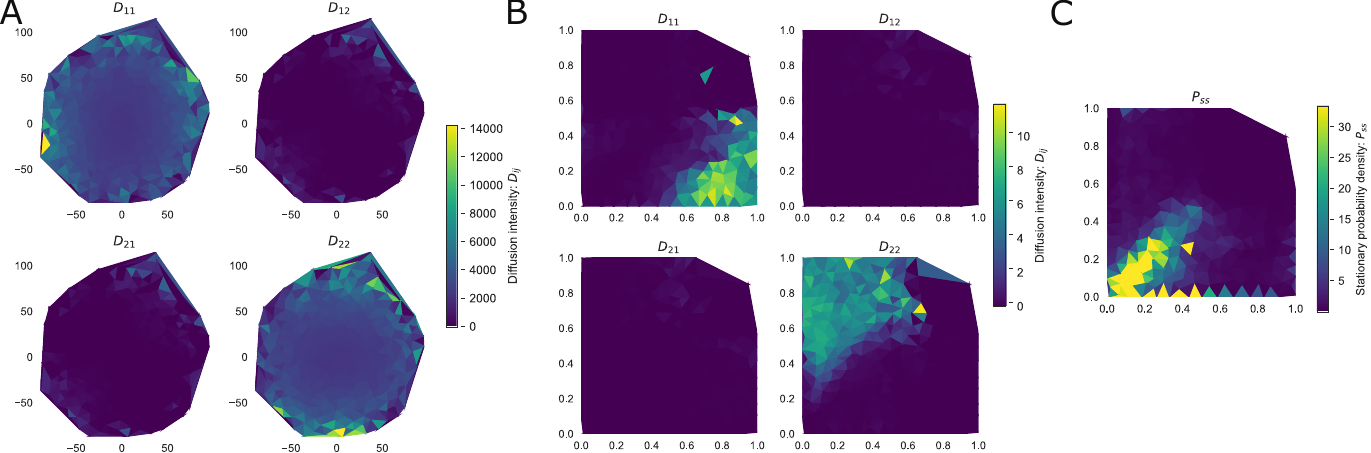}
    \caption{\textbf{Diffusion estimation in real-world data.} \textbf{A.} An example of the spatially-dependent diffusion estimator in the RBC `Healthy 1'. \textbf{B.} An example of the spatially-dependent diffusion estimator in the healthy heartbeats. \textbf{C.} A maximum-likelihood estimation of the stationary probability distribution $P_{ss}$ in the healthy heartbeats.}
    \label{fig: realworld diffusion}
\end{figure*}
The framework presented here assumes that diffusion is not spatially-dependent and thus can be represented by a single matrix $\bm{D}$. Here, we briefly assess whether this is the case in the real-world data by leveraging the spatially-dependent estimator $\bm{\hat{D}}$ of Eq. \ref{eq: diffusion}. Panel A of Fig. \ref{fig: realworld diffusion} shows the coefficients of the diffusion estimator in each of the triangular bins, estimated using the RBC `Healthy 1'. The diffusion appears to be spatially-independent, with more variation at the borders where the trajectories have fewer points leading to higher estimation error. Moreover, the diffusion is close to isotropic. Panel B shows the diffusion estimator calculated on the healthy heartbeats. This appears to show more spatial dependence whilst remaining mostly isotropic. Nevertheless, by estimating the stationary distribution with maximum-likelihood estimation, shown in panel C, we can see that the trajectories spend relatively little time in the parts of phase-space that diffusion variation which indicates that this variation could arise from the numerical error associated with having only a few samples in this area of phase-space rather than a genuine spatial dependence.
\subsubsection{Ensemble bootstrapping}
\label{app: data bootstrap}
The finite length of the trajectories limits the accuracy to which the force fields can be reconstructed and decomposed. Furthermore, as our approach ingests the ensemble of trajectories in order to reconstruct flows, we obtain a single number for each condition. In order to estimate these errors and thus to perform significance testing, we employ ensemble bootstrapping \cite{battle2016brokendetailedbalance,Lynn2021brokendetailedbalance}. Each ensemble is of the form $\{\mathbf{x}_{t_1,r},...,\mathbf{x}_{t_T,r}\}_{r=1}^{R}$. We construct a bootstrapped ensemble by sampling trajectories from the original ensemble with replacement. We use the same number of trajectories in each ensemble and use 100 bootstraps resulting in 101 ensembles in each condition. As shown in Fig. \ref{fig: Data}, this gives a distribution of $P_c$, upon which we perform independent $t$-tests for significance.
\subsubsection{Surrogate-testing with shuffled data}
\label{app: data shuffling}
\begin{figure}
    \centering
    \includegraphics[width=\linewidth]{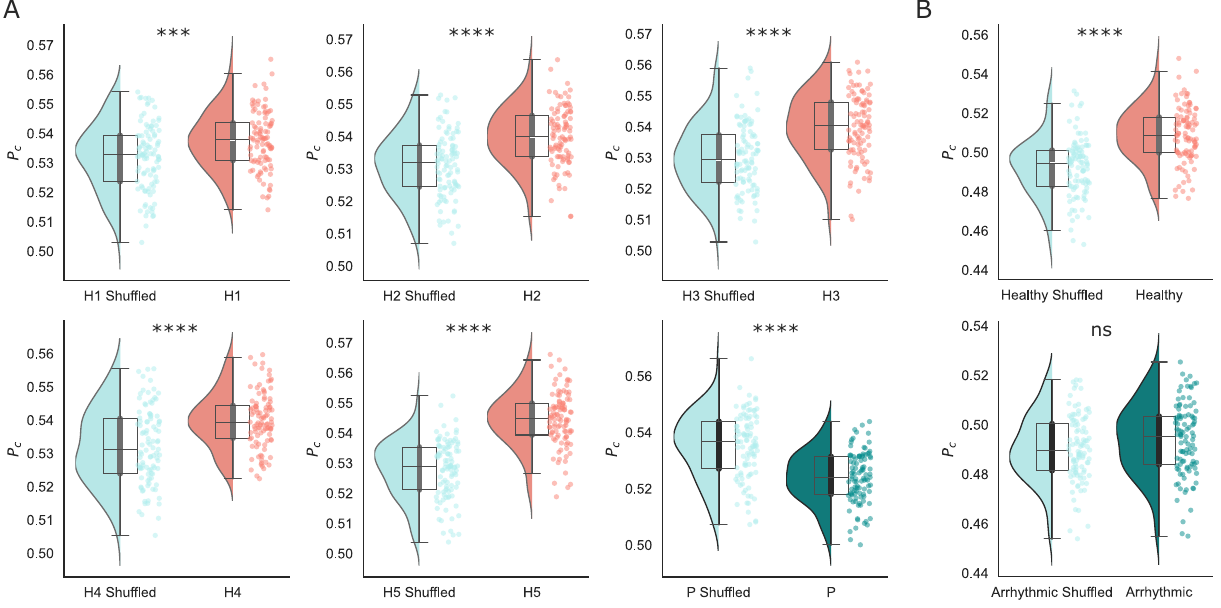}
    \caption{\textbf{Surrogate-testing with shuffled time-series}: For each condition in either dataset, we construct 100 bootstraps. For each bootstrap, we shuffle each of the trajectories randomly in time, apply the method, and calculate the proportion of flow in the curl space for both the original and shuffled data. We then run paired $t-$tests to check if the proportion of curl flow is significantly higher or lower than a random sample. Panel A shows the result for the five healthy and one passive RBC. Panel B shows the result for the healthy and arrhythmic heartbeats. ($p<0.001$,***), ($p<0.0001$, ****) and ($p>0.05$, ns).}
    \label{fig: shuffling}
\end{figure}
In order to assess whether the proportion of curl-flow measured is due to the statistical structure of the time-series or due to a combination of chance and the finite-length of the data, we compare our time-series to surrogate `null' data. In order to break the temporal ordering of the time-series, we shuffle randomly over the time-points. This is in contrast with other typical randomisation techniques, such as phase-randomisation, which do not disrupt the temporal ordering sufficiently \cite{Lynn2021brokendetailedbalance}. For each condition in each dataset, we construct a 100 bootstrapped ensembles of delay-embedded trajectories. We then shuffle the temporal ordering of the points in phase-space randomly in time. Next we apply our approach to both the original and shuffled ensemble to calculate two measures of $P_c$. We then perform paired $t-$tests to identify if the statistical structure of the time-series results in higher or lower $P_c$ than could be expected by chance. In panel A of Fig. \ref{fig: shuffling},  we present the results for the RBCs and in panel B we present the results for the healthy and arrhythmic heartbeats. We find that all healthy RBCs have significantly higher $P_c$ that the shuffled time-series indicating a high prevalence of curling flow, whilst the passive cell have significantly lower $P_c$ than the shuffled time-series indicating a very lower prevalence of curling flow. Similarly, for the healthy heartbeats, we find a significantly high level of $P_c$ compared to shuffled data, whilst for arrhythmic heartbeats, we find that $P_c$ is comparable between the null data and the original series.
\subsubsection{Variation of the time-delay parameter}
\label{app: data timedelayvariation}
\begin{figure}
    \centering
    \includegraphics[width=\linewidth]{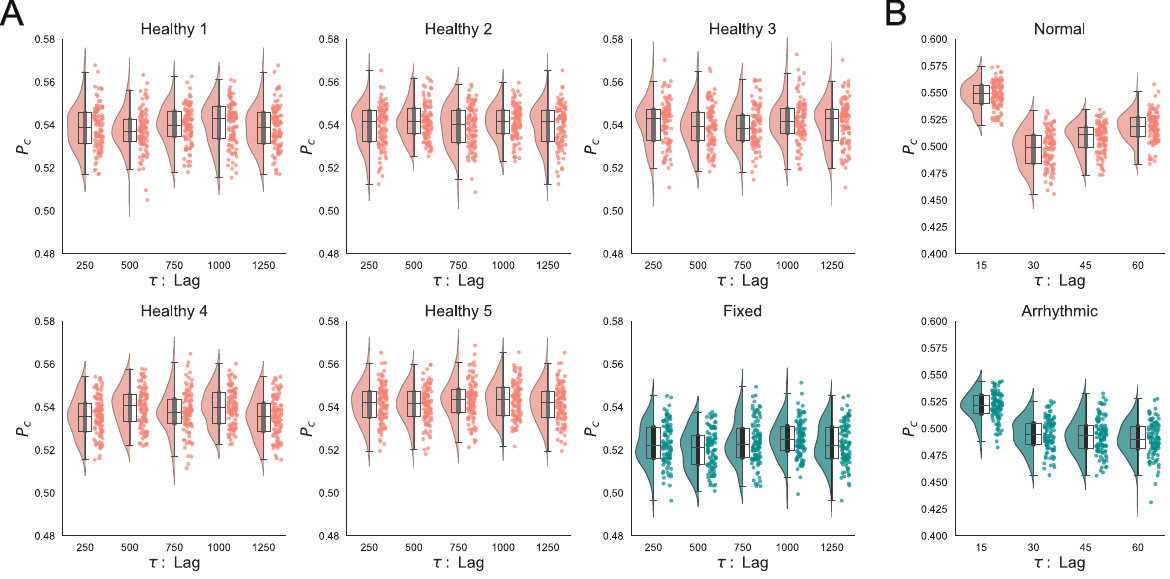}
    \caption{\textbf{Variation of the lag parameter, $\tau$, in the delay embedding.} A. Varying the lag parameter in the RBC experiments shows little to no change suggesting that the phase-space dynamics are comparable for a range of different lags. B. In the case of the heartbeat dynamics, the time-series are much shorter as a result, varying the lag parameter causes variation in the quantity, but the relative relationship between normal and arrhythmic heartbeats is preserved.}
    \label{fig: lag variation}
\end{figure}

\Rev{When analysing the real-world data, we used a single lag for each experiment that was chosen to be the mean first-zero-crossing of the autocorrelation function, which is a conventional choice \cite{kantz2003nonlinear}. For the RBC experiments we obtain $\tau = 998, (T=10,000)$, whilst for the heartbeats we obtain $\tau = 45, (T=188)$. In Fig. \ref{fig: lag variation}, we consider the effect of varying this parameter. For the RBCs, we consider lags of 250, 500, 750, 1000 and 1250. For the heartbeats, we consider lags of 15, 30, 45 and 60. For the RBCs, panel A shows that variation in the lag parameter shows no significant change in the distribution of $P_c$ over 100 boot-strapped samples, whilst the relative relationship between healthy and passive/fixed cells is kept the same. In panel B, we can see that variation in the lag parameter can cause greater variation in $P_c$ for the heartbeat dynamics. This is because the heartbeat data has a relatively short-time series which is made even shorter by an increase in the lag. This can increase the error in the measurements and lead to greater variation. Nevertheless, the qualitative relationship between normal and arrhythmic heartbeats appears to hold. This analysis suggests that our approach is not very sensitive to changes in the lag parameter provided the lag is relatively small in comparison with the length of the time-series.} 
\subsubsection{Classifying time-series with standard methods}
\label{sec: benchmark}
\begin{figure}
    \centering
    \includegraphics[width=\linewidth]{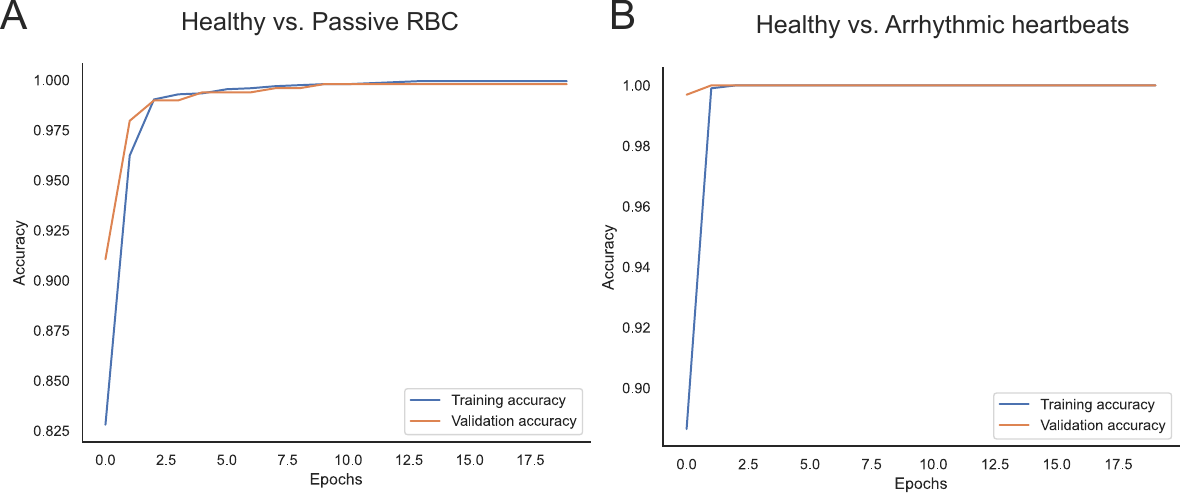}
    \caption{\textbf{Classifying recordings with a neural network}: We attempt to classify the time-series data considered here with a neural network trained on the summary statistics of the data. We find that for both the RBCs and the heartbeats, the neural network achieves almost perfect classification after very little training. This is due to clear differences in the summary statistics that allow for easy classification.}
    \label{fig: ML}
\end{figure}
\begin{figure}
    \centering
    \includegraphics[width=\linewidth]{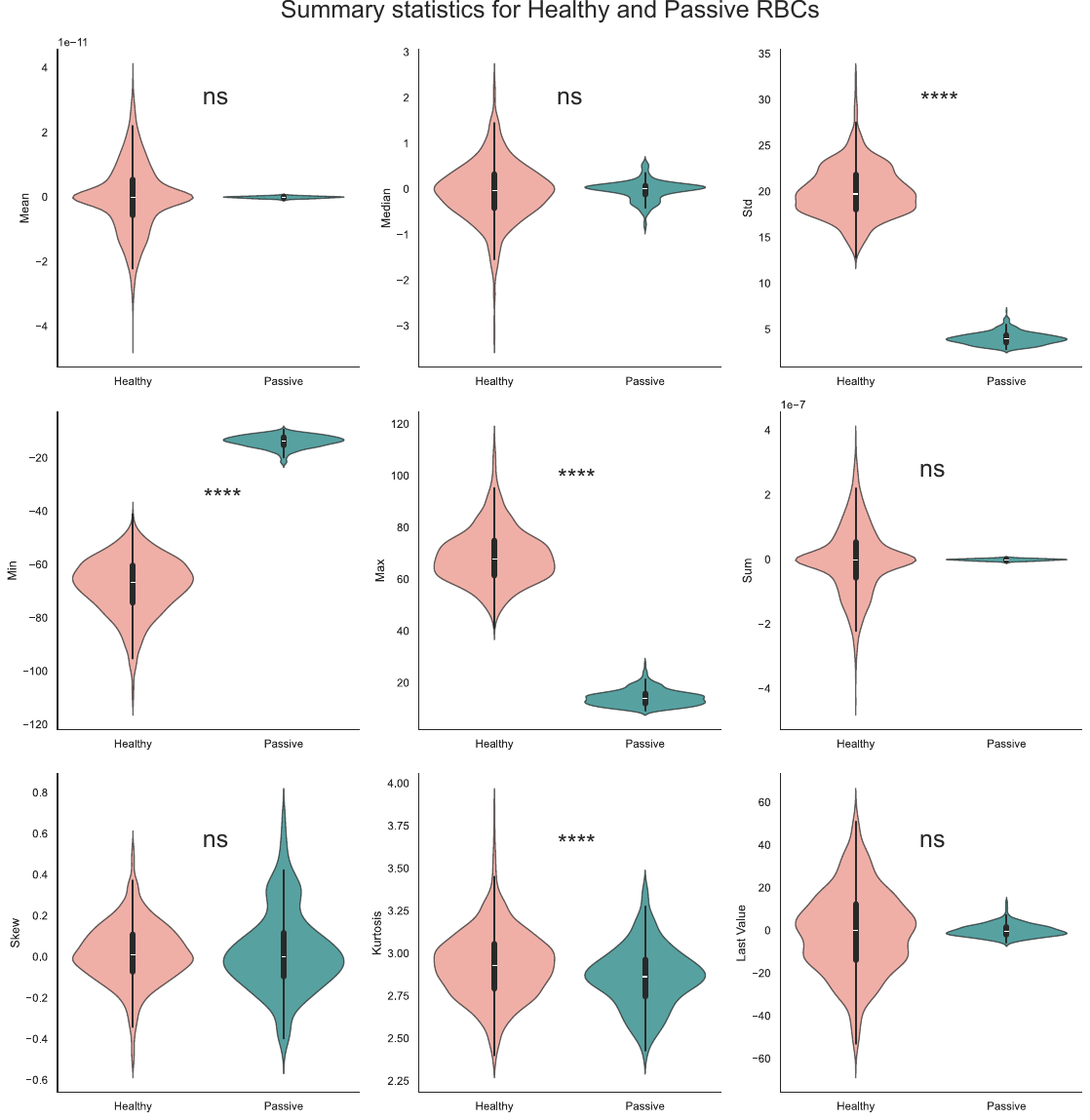}
    \caption{\textbf{Distribution of summary statistics for normalised red-blood cell data}: We consider 10 summary statistics as features to input into the neural network. Here we show that some statistics, including standard deviation, Min, Max and kurtosis, are significantly different between healthy and passive cells, making for easy classification of the recordings. Significance is denotes as (ns) if $p>0.05$, (*) if $p<0.05$, (**) if $p<0.01$, (***) if $p<0.001$ and (****) if $p<0.0001$.}
    \label{fig: summarystatsRBC}
\end{figure}
\begin{figure}
    \centering
    \includegraphics[width=\linewidth]{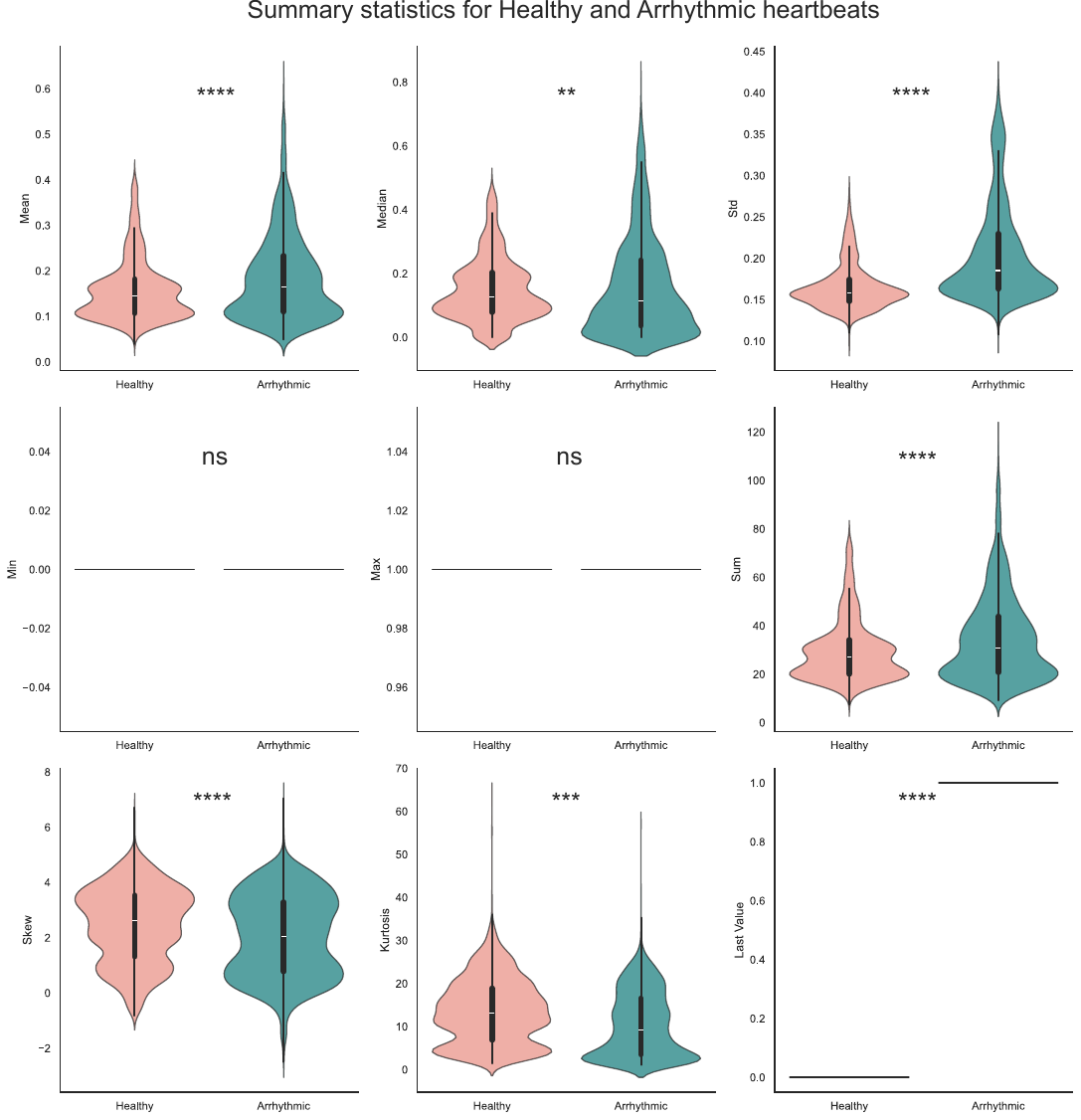}
    \caption{\textbf{Distribution of summary statistics for normalised heartbeat data data}: We consider 10 summary statistics as features to input into the neural network. Here we show that some statistics, including mean, median, standard deviation, sum, skew, kurtosis, and Last Value are significantly different between healthy and arrhythmic heartbeats, making for easy classification of the recordings. Significance is denotes as (ns) if $p>0.05$, (*) if $p<0.05$, (**) if $p<0.01$, (***) if $p<0.001$ and (****) if $p<0.0001$.}
    \label{fig: summarystatsHeart}
\end{figure}
\begin{figure}
    \centering
    \includegraphics[width=0.5\linewidth]{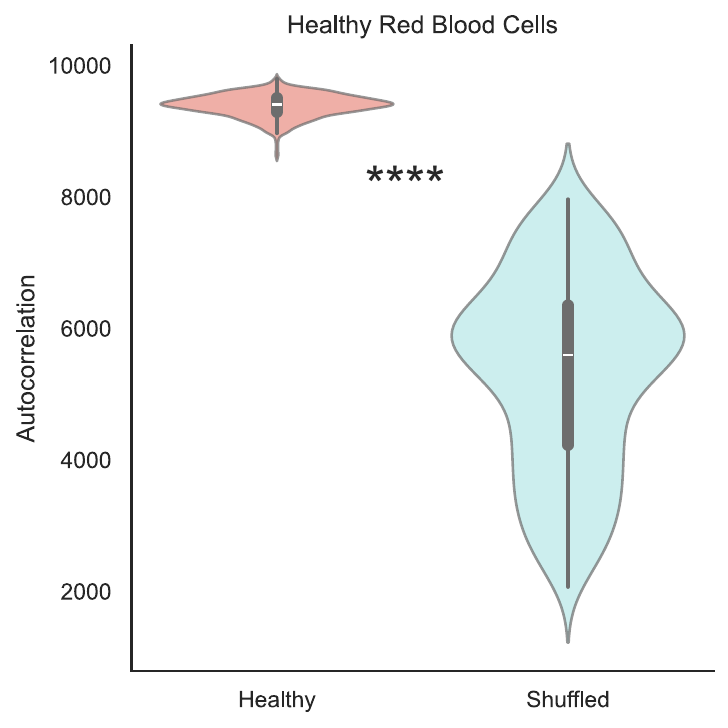}
    \caption{\textbf{Distribution of auto-correlation for healthy RBC data and shuffled surrogate data}: One can distinguish between true dynamics and shuffled surrogate data using temporally-dependent summary statistics. Here we show that the auto-correlation of data from the RBC Healthy 1 is significantly higher than its shuffled surrogate. Significance is denotes as (ns) if $p>0.05$, (*) if $p<0.05$, (**) if $p<0.01$, (***) if $p<0.001$ and (****) if $p<0.0001$.}
    \label{fig: shuffledRBC}
\end{figure}

\Rev{Whilst we have shown that our method can differentiate between recordings from biological systems in different conditions, our main goal was to gain insight into the underlying reversible and irreversible dynamics directly from stochastic trajectories. Our approach allows for insight into the thermodynamics and nonequilibrium physics of the biological processes, and goes beyond a classification tool \cite{fang2019nonequilibrium}. Measuring $P_c$, the proportion of the flow in the curl space, is a first application that allows us to validate that our approach does capture irreversible flows by confirming results from the literature \cite{peng2019nonequilibrium,costa2005heart}. We note that differentiating between conditions using the time-series data considered here is not a challenging task in itself, as we will show in this section. Beyond the prevalence of irreversible flows, the underlying time-series have significant differences in their statistics, that allow for easy classification. To illustrate this, we consider a neural-network classifier and show that it achieves (almost) perfect classification of recordings when trained on simple time-series summary statistics. For each of recording in the dataset, we compute a battery of trajectory features in the form of summary statistics. In particular, we consider the mean, median, standard deviation (Std), minimum value (Min), maximum value (Max), sum, skew, kurtosis and last value of each signal. For the RBCs, recordings are either healthy or passive, whilst for heartbeats, recordings are either normal or arrhythmic. We then train a neural-network on the summary statistics of a subset of the recordings with the goal of classifying each recording into its respective category. Our network has 2 hidden layers with 32 and 16 `ReLu' units respectively and an output layer with sigmoidal activation. We use a binary cross-entropy loss function and the Adam optimiser. We train the network for 20 epochs with a validation split of 0.2 and a batch size of 50. In addition, we scale all recordings between [0,1] before training and use class-weighting to account for the different densities of the classes in the dataset. As shown in Fig. \ref{fig: ML}, we find that the neural network has (almost) perfect accuracy after only a few epochs of training. On the test set, we find that it classified both RBC and heartbeat recordings with 100\% accuracy. Fig. \ref{fig: summarystatsRBC} shows the distribution of the summary statistics, after data normalisation, for both healthy and passive cells. A subset of the statistics, namely the standard deviation, Min, Max and kurtosis are significantly different between healthy and passive cells, therefore, these features allow for simple classification of trajectories. Similarly, Fig. \ref{fig: summarystatsHeart} shows the corresponding summary statistics for healthy and arrhythmic heartbeats, where the mean, median, standard deviation, sum, skew and kurtosis are all found to be significantly different.\\
\\
Our approach is also able to distinguish between nonequilibrium dynamics such as those found in healthy RBCs, and the equilibrium dynamics of surrogate data generated by shuffling recordings in time. All the summary statistics considered in the previous experiment are invariant under shuffling of the data in time, which suggests that such trajectory features are insufficient for such differentiation. However, one can select summary statistics that depend on the temporal ordering, such as auto-correlation, to distinguish true and shuffled recordings. In Fig. \ref{fig: shuffledRBC}, we show that the auto-correlation of data from Healthy 1, the RBC, is significantly higher than a surrogate generated by shuffling the data in time. This allows for the easy distinguishing of true dynamics from shuffled counterparts based on simple summary statistics. 
}
\subsubsection{Flickering experiments with red-blood cells}
\label{app: data rbc}
RBCs metabolise glucose causing the cell membrane to actively \textit{flicker} with the dissipated energy and the created entropy \cite{turlier2019unveiling,betz2008rbc,diterlizzi2024variancesum}. By depleting the ATP of a healthy RBC, one obtains a \textit{passive} cell whose flickering is closer to reversible thermal noise \cite{rodriguezgarcia2015directcytoskeleton,diterlizzi2024variancesum}. The position of the RBC membrane is captured using optical microscopy to form movies at a sampling rate of 2000 frames per second. The position of a point on the membrane is extracted to give position (in nm) traces with 10,000 time-points. The dataset includes 512 position traces from each of five healthy cells and one passive cell. Further discussion of the data can be found in Ref. \cite{rodriguezgarcia2015directcytoskeleton,diterlizzi2024variancesum} and the data is openly available at Ref. \cite{diterlizzi2024data}.
\subsubsection{Healthy and arrhythmic heartbeats}
\label{app: data heart}
Human heartbeats can be recorded using an ECG, which measures the voltage of electrical signals in the heart over time. Previous research has shown that the time-irreversibility of these signals is disrupted during aging and disease \cite{costa2005heart}. The dataset considered here is the PTB Diagnostic ECG Database \cite{bousseljot1995cardiodat,goldberger2000physionet} as pre-processed in \cite{kachuee2018heart}. The dataset contains time-series for individual extracted beats with either `normal' (healthy volunteers) or `abnormal' (myocardial infarction) dynamics. The dataset contains 4046 normal heartbeats and 10,506 abnormal heartbeats each composed of 188 time-points at a sampling frequency of 125Hz. We randomly sampled 4046 of the abnormal heartbeats for our analysis. The data is openly available at Ref. \cite{fazeli2018heartdata}.
\subsection{The Helmholtz-Hodge decomposition of the Ornstein-Uhlenbeck process}
\label{app: OU}
The OU process is given by,
\begin{align}
    \frac{d\mathbf{x}}{dt}&=-\bm{B}\mathbf{x}(t)+ \xi(t),
\end{align}
with $\langle\bm{\xi}(t),\bm{\xi}(t')\rangle = 2\bm{D}\delta(t-t')$. If the real-part of the eigenvalues of $\bm{B}$ are all positive, then the process will converge to a Gaussian stationary distribution with mean zero and covariance satisfying \cite{Godreche2018OU},
\begin{align}
\bm{S} &= \lim_{t\rightarrow \infty}[ e^{\bm{-B}t}\bm{S}(0)e^{\bm{-B}^{\top}t} 
+ 2\int_0^{t} e^{\bm{-B}(t-s)}\bm{D}e^{\bm{-B}^{\top}(t-s)}\;ds],\\
&= 2\int_0^{\infty} e^{\bm{-B}t}\bm{D}e^{\bm{-B}^{\top}t}\;dt.
\end{align}
It can also be shown that $\bm{S}$ satisfies the following Sylvester equation,
\begin{align}
\label{eq: sylvester}
\bm{B}\bm{S} + \bm{S}\bm{B}^{\top}&=2\bm{D}.
\end{align}
Next, we define the Onsager matrix, $\bm{L}$, of kinetic coefficients,
\begin{align}
\bm{L}&=\bm{BS}=\bm{D}+\bm{Q},\\
\bm{L}^{\top}&=\bm{SB}^{\top}=\bm{D}-\bm{Q},
\end{align}
which parameterising the asymmetries through the anti-symmetric matrix $\bm{Q}$ which captures the rotational component of $\bm{B}$ \cite{Godreche2018OU}. Therefore the HHD of the OU process is given by,
\begin{align}
    \bm{B}\mathbf{x}&= (\bm{B}_{\text{irrev}}+\bm{B}_{\text{rev}})\mathbf{x},\\
    \bm{B}_{\text{irrev}}&= \bm{QS}^{-1},\\
    \bm{B}_{\text{rev}}&= \bm{DS}^{-1},
\end{align}
where the matrix $\bm{B}_{\text{rev}}$ is symmetric. Additionally, the stationary probability fluxes are given by, $\bm{J} = -\bm{QS}^{-1}\mathbf{x}$ \cite{Godreche2018OU}, used to display the arrows in panel B of Fig. \ref{fig: Helmholtz}.

\end{document}